\documentclass[10pt,twocolumn]{revtex4-1}
\usepackage{bm}
\usepackage{hyperref}
\usepackage{makeidx}
\usepackage{amsmath}
\usepackage{graphicx}
\usepackage{dcolumn}
\usepackage{color}
\usepackage{graphics}
\usepackage{amsfonts}
\usepackage{amssymb}
\usepackage{epsfig}
\usepackage{float}
\usepackage[british,UKenglish,USenglish,english,american]{babel}
\usepackage{epstopdf}
\usepackage{graphics, subfigure}
\usepackage{array}

\begin{document}

\title{ Effects of anisotropic correlations in fermionic zero-energy bound states of topological phases }
\author{ M. A. R. \surname{Griffith} }
\email{griffithphys@gmail.com}
\author{ E. \surname{Mamani} }
\author{ L. \surname{Nunes} }
\author{ H. \surname{Caldas} }

\affiliation{Universidade Federal de S\~{a}o Jo\~{a}o del-Rei, \\Rua , 22290-180,   S\~{a}o Jo\~{a}o del-Rei, MG, Brazil}

\date{\today }

\begin{abstract}
Topological phases of matter have been used as a fertile realm of intensive discussions about fermionic fractionalization. In this work, we study the effects of anisotropic superconducting correlations in the fermionic fractionalization on the topological phases. We consider a hybrid version of the SSH and Kitaev models with an anisotropic superconducting order parameter to investigate the unusual states with zero energy that emerges in a finite chain. To obtain these zero energy solutions, we built a chain with a well-defined domain wall at the middle of the chain. Our solutions indicates an interesting dynamic between the zero-energy state around the domain wall and the superconducting correlation parameters. Finally, we find that the presence of an isolated Majorana at the ends of the chain is strongly dependent  on the existence of the solitonic excitation at the middle of the chain.
\end{abstract}

\maketitle

\section{Introduction}

In condensed matter physics, Majorana zero-energy bound states are quasi-particle excitations that arise when a single electronic mode fractionalize into two halves \cite{Kitaev}. The Kitaev chain has been used to study the properties of non-local Majorana zero-energy bound states that reside at edges of the chain \cite{Kitaev, Alicea, Leijnse,Lutchyn,Oreg,Sato,Potter,Beenakker, Motohiko, Ning,Bernevig}.

The Kitaev model, which describes an one-dimensional (1D) spinless $p$-wave superconductor chain, belongs to the BDI topological class \cite{Bernevig}. In this class, the non-trivial topological phase depends on time-reversal symmetry, particle-hole symmetry and chiral symmetry \cite{Altland, Schnyder}. The special particle-hole symmetry ensures that the Majorana zero-energy bound states behave like a superposition of particle and hole and have no well-defined charge \cite{Kitaev,Bernevig}. However, besides the particle-hole symmetry, the number of pairs of zero-energy bound states at the edges is crucial to obtain one isolated Majorana zero-energy bound state. Any chain with an even number of zero-energy bound states per edge does not have an isolated Majorana at the edges. To obtain an isolated Majorana zero-energy bound state is necessary an odd number of zero energy at the ends of the chain \cite{Bernevig}.

In the last years, several models have been proposed to support isolated Majorana zero-energy bound states \cite{Motohiko,Ning,Sticlet,Wakatsuki,Yahyavi}. An interesting situation happens when the superconducting chain is subjected to dimerization. Recently, R. Wakatsuki \emph{et al} \cite{Wakatsuki} proposed a tight-binding model for the investigation of this {\it dimerized Kitaev model}, which has been investigated by several authors as, for example, in \cite{Xiong,Wang,Ezawa2,Hua}.

The SSH-like ground state displays an even number of zero-energy states, while the Kitaev-like ground states present an odd number. In the Kitaev-like ground states, the model can support an isolated Majorana zero-energy bound states at the ends of the chain \cite{Wakatsuki}.

Therefore, the dimerized Kitaev chain allows to study topological phase transitions from ground states with an even number of zero-energy states to the ground state with an odd number of zero-energy states. Considering a domain wall at the middle of the chain, the authors of Ref~\cite{Wakatsuki} studied the evolution of the solitonic mode across the topological phase transition from SSH-like ground state to  Kitaev-like ground state. The solitonic mode suffers a split across the quantum phase transition between the SSH-like and Kitaev-like phases. The splitting of the solitonic mode gives rise to one Majorana zero-energy bound state at each end of the chain. We point out that, the results from Ref~\cite{Wakatsuki} depend on a specific constraint between the hopping terms and superconducting correlation (i.e., a constraint between the hopping and superconducting parameters such that they are interdependent).

Motivated by the recent experimental observation of spectroscopic signatures of Majorana zero-energy modes in semiconducting nanowires placed on the surface of superconducting substrate \cite{Mourik,Nadj}, in this work we explore the phase diagram of the Kitaev model with alternating hoppings and superconducting correlations.

Referring specifically to the dimerized Kitaev chain, we could cite the following proposals as possible experimental achievements:

In reference \cite{Tadeu} the authors suggested that a nanowire with strong spin-orbit interaction could be deposited on a s-wave superconducting substrate. The effects of misalignment between the atoms of nanowire and the substrate lattice can promote the desirable hopping dimerization in the nanowire. After reaching the dimerization, one can turn on a magnetic field along the axial axes of the nanowire and adjust the chemical potential to obtain a quasi-particle spectrum of a spin-triplet superconductor in the nanowire, as happens in the case of the usual p-wave Kitaev model \cite{Kitaev}.
We believe that this method is very promising and realistic because the interaction between the nanowire and substrate, which induces the dimerization, can indeed be realized by a type of proximity effect \cite{Lutchyn}.

The second possibility comes from a recent experimental realization in artificial lattice vacancies \cite{Drost}. The authors were able to obtain topological states in engineered atomic lattices. They showed that it is possible to build a dimerized chain through vacancies on the chlorine monolayer on a $Cu(100)$ surface.  The essential physics of this topological systems can be captured by tight-binding models. Besides, this type of lattice provides an excellent platform to study the domain wall and the states that arise around it. Therefore, our analyses could be used to investigate topological states in vacancy latices in the presence of correlated electronic modes.

Finally, the dimerized Kitaev chain could, in principle, be done in an ``artificial''  one-dimensional lattice, comprised of an array of trapped ultra cold atoms with effective hoppings, as has been realized recently in a bosonic version of the Su-Schrieffer-Heeger (SSH) model\cite{Leseleuc}.

As we have anticipated, our results do not depend on constraints between the hopping terms and superconducting parameters, as employed in Ref~\cite{Wakatsuki}.
The effects of the anisotropic correlations have been clarified. We provide the correct phase transition between the hybrid model and the pure SSH model.
We studied the effects of the anisotropic hopping terms and anisotropic superconducting correlations over the zero energy bound state around the domain wall at the middle of the hybrid chain. To this purpose, we create a general kink (i.e., in hopping and superconducting correlations) at the middle of the chain.

The paper is organized as follows, in section \ref{sec2}, we defined the model and its important discrete symmetries.
In Section \ref{sec3}, we introduce the topological invariants related to each symmetry and the phase diagrams of the system.
In Section \ref{sec4}, we simulate a kink at the middle of the chain to obtain the zero-energy states around the domain wall. We investigated the effects of the anisotropic superconducting correlation over the solitonic model. Finally, we summarize our results in the conclusion Section.

\section{ Anisotropic Model} \label{sec2}

\begin{figure}[t!]
  \centering
  \includegraphics[width=1\columnwidth]{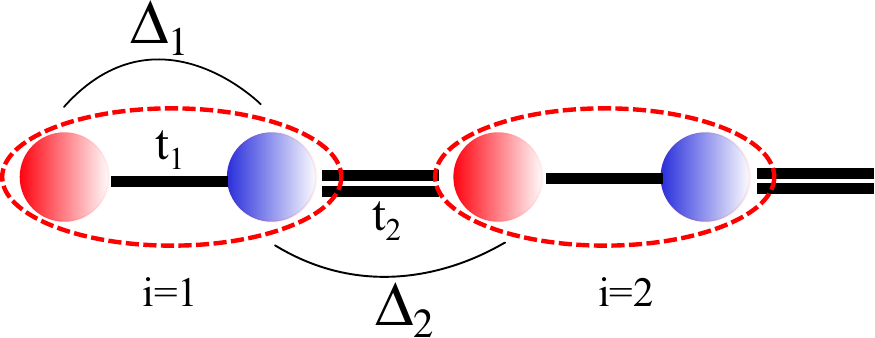}
  \caption{(Color online) The hybrid  model with nearest neighbors $t_1$ and $t_2$ hopping and superconducting parameters $\Delta_1$ and $\Delta_2$. Each unit cell $i$ contains a pair of sublattice A and B, the red and blue spheres, respectively. The hopping terms and superconducting parameters works for different sublattices, $t_{1}$ and $\Delta_1$ within the unit cell while $t_{2}$ and  $\Delta_{2}$  out of the unit cell.}\label{SSHFig}
\end{figure}

The SSH model, \ref{SSHFig}, subjected to anisotropic superconducting correlations, can be defined as

\begin{eqnarray}\label{Hibrido}
  H_{Hyb}&=& \sum_i
  \left(
    t_1 c^{\dagger}_{Bi} c_{A,i}
  + t_2 c^{\dagger}_{A i+1} c_{B,i}
  \right)
  + H.c.
  \\ \nonumber
   &-&
   \sum_i
   \left(
   \Delta_{1} \, c^{\dagger}_{Bi} c^{\dagger}_{Ai}
   -
   \Delta_{1} c^{\dagger}_{A,i}, c^{\dagger}_{B,i} + H.c.
   \right) \\ \nonumber
   &-&
   \sum_i
   \left(
   \Delta_{2} \, c^{\dagger}_{Ai+1} c^{\dagger}_{Bi}
   -  \Delta_{2} c^{\dagger}_{B,i}, c^{\dagger}_{A,i+1} + H.c.
   \right)
   \\ \nonumber
   &-&
   \mu \sum_{ i, \alpha } c^{\dagger}_{\alpha,i} c_{\alpha,i }
   %%%%%%%%%%%%%%%%%%%%%%%%
   \, ,
\end{eqnarray}
where $\mu$ is the chemical potential, $t_1$ and $t_2$ are hopping terms and $\Delta_1$ and $\Delta_2$ are the superconductors anisotropic correlations order parameters, see Fig.~\ref{SSHFig}. Here, $c_{\alpha,i}$ is an operator that creates a fermion at the site $i$ in the sublattice $\alpha$ with $\alpha =A,B$. The model described by Eq.~\ref{Hibrido} belong to the BDI topological class in the Altland-Zirnbauer classification of topological superconductors/insulators \cite{Altland,Schnyder}. The model possesses time-reversal symmetry, particle-hole symmetry and chiral symmetry or sublattice symmetry \cite{Wakatsuki}.

The Hamiltonian in Eq.~(\ref{Hibrido}) can be rewritten in $k$ space in terms of the Dirac matrices as

\begin{eqnarray}\label{Hibrido2}
  H_{Hyb} =
  \sum_k \psi_k^\dagger \mathcal{ H }_k \psi_k \, ,
\end{eqnarray}
where
$
\psi^{\dagger}
=
\left(
c^{\dagger}_{A,k},c^{\dagger}_{B,k},c_{A,-k},c_{B,-k}
\right)
$ and

\begin{eqnarray}\label{Hk}
\mathcal{ H }_k=&-&\mu \sigma_3 \otimes \sigma_0 + a \sigma_3 \otimes \sigma_1 + b \sigma_3 \otimes \sigma_2 \\ \nonumber
   &-& c \sigma_2 \otimes \sigma_2 - d \sigma_2 \otimes \sigma_1
\end{eqnarray}
with
$ a = t_1+t_2 \cos(ka) $,
$ b = t_2 \sin(ka) $,
$ c = \Delta_1+\Delta_2 \cos(ka)$ and
$ d = \Delta_2 \sin(ka) $.

When $\Delta_1=\Delta_2=\mu=0$ the system becomes the SSH model that belongs to the same topological class as the Kitaev model, as is well known \cite{Wakatsuki,Bernevig}.

The SSH model does not display Majorana zero-energy bound states at the ends of the chain \cite{Bernevig}.

In non-trivial topological phase ($|t_2|>|t_1|$), the SSH model exhibits one pair of zero-energy states. These zero-energy states are spread over
several sites near the edge of the chain. These two zero-energy edge states are localized around the first and the last site, respectively.
The degree of the localization of these states depends on the distance from the critical point $t_2=t_1$, or more precisely,
its localization depends on the penetration depth $\xi$ and the critical exponent $\nu$. For example, for SSH model $\xi=(t_2-t_1)^{-\nu}$,
where $t_2=t_1$ is the critical point and $\nu=1$ is the critical correlation exponent of the SSH model. Indeed, for the case of the simple SSH, these zero-energy state can be written as $\Psi(n) \propto e^{-n /\xi}$ \cite{Rufo}. Note that away from the critical point, these zero-energy states are majority localized around the first and last sites of the chain.

The zero-energy states of the SSH model are composed by superposition of particles and exhibit a well-defined charge. Therefore, a conventional fermionic excitation mode emerges \cite{Bernevig}.

In the SSH model, the chiral symmetry results from an equivalence between the sublattices A and B.  We notice that the chiral symmetry protects the non-trivial topological region and the zero-energy edge states \cite{Bernevig}.

Now, when $t_1=t_2$ and $\Delta_1=\Delta_2$, the Hamiltonian, Eq.~\ref{Hibrido}, becomes the Kitaev model, which possesses particle-hole symmetry, given  ($\Xi=\sigma_1 \mathcal{K}$), time reversal symmetry ($\Theta=\mathcal{K}$) and  also chiral symmetry ($\Pi = \Xi \Theta$) \cite{Wakatsuki}.

Nevertheless, although the Kitaev model and the SSH model belongs to the same topological classification (i.e. BDI), it is important to notice that the origin of the particle-hole symmetry is physically distinct for each model \cite{Bernevig,Wakatsuki}.

As pointed out in Ref~\cite{Bernevig}, for the SSH model, the particle-hole symmetry was induced by time-reversal and chiral symmetries, while for the Kitaev model, the particle-hole symmetry is essentially a characteristic of the superconducting phase.

\section{Phase Diagram on the parameter space }\label{sec3}

\begin{figure}[h!]
  \centering
  \includegraphics[width=1\columnwidth]{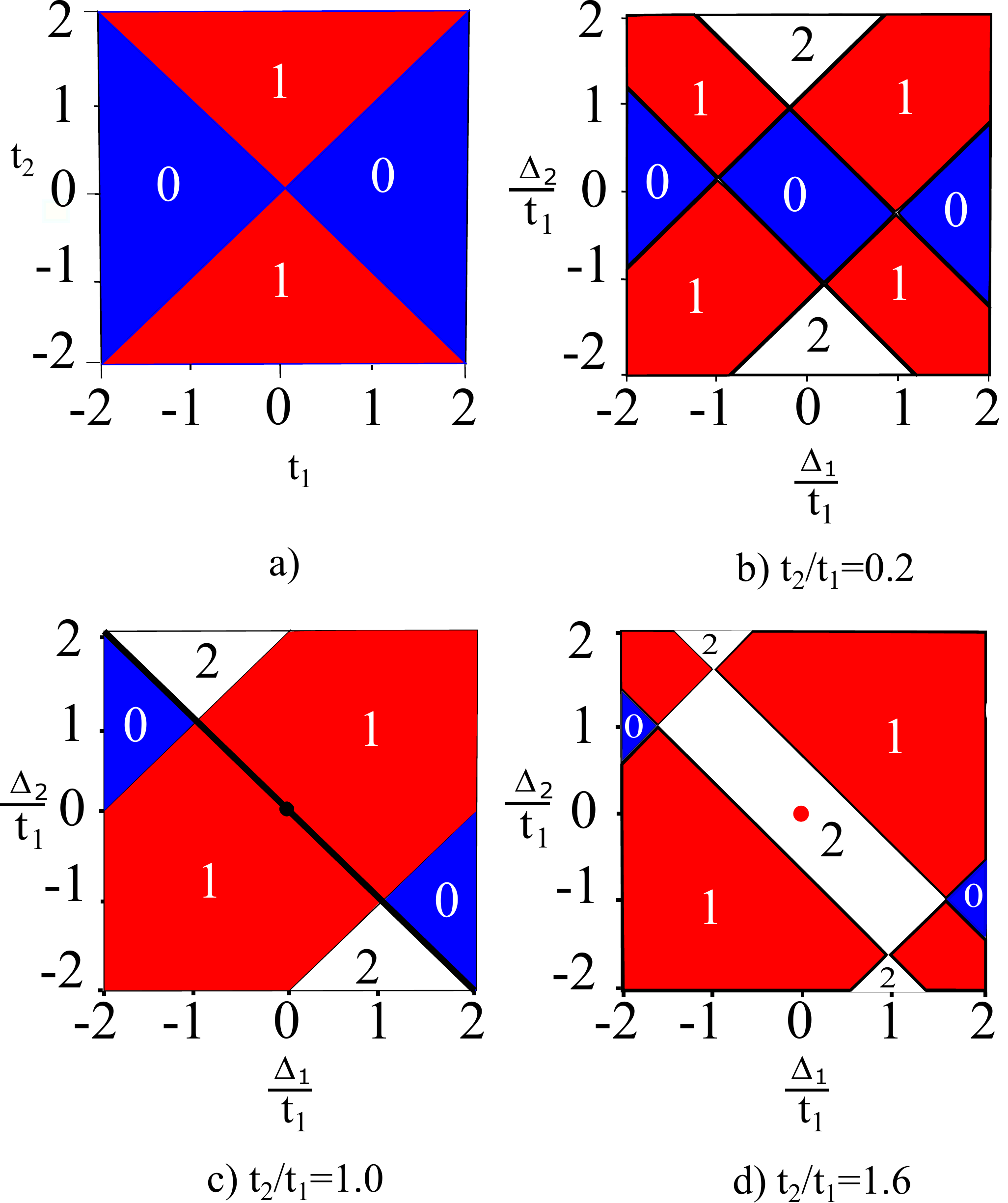}
  \caption{(Color online) Topological phase diagrams with respect to $W_1$ with $\mu=0$. The numbers in the figures (a), (b), (c) and (d) denotes $W_1$.
  (a) Topological phase diagram of the pure SSH model with $\Delta_1=\Delta_2=0$. The trivial topological region is  indicated by blue color ($W_1=0$), while the red region is the non-trivial topological phase with $W_1=1$. In figures (b), (c) and (d), we considered the effects of the correlations $\Delta_i$ for fixed values of the ratio $t_2/t_1$.  (b) Topological phase diagram with $\frac{t_2}{t_1}=0.2$, (c) Topological phase diagram with $\frac{t_2}{t_1}=1$ and (d) Topological phase diagram with $\frac{t_2}{t_1}=1.6$. A new topological phase with $W_1=2$ (white regions) arises in the superconducting hybrid system. }\label{DiagramaMuzero}
\end{figure}

\begin{figure}[t!]
  \centering
  \includegraphics[width=1\columnwidth]{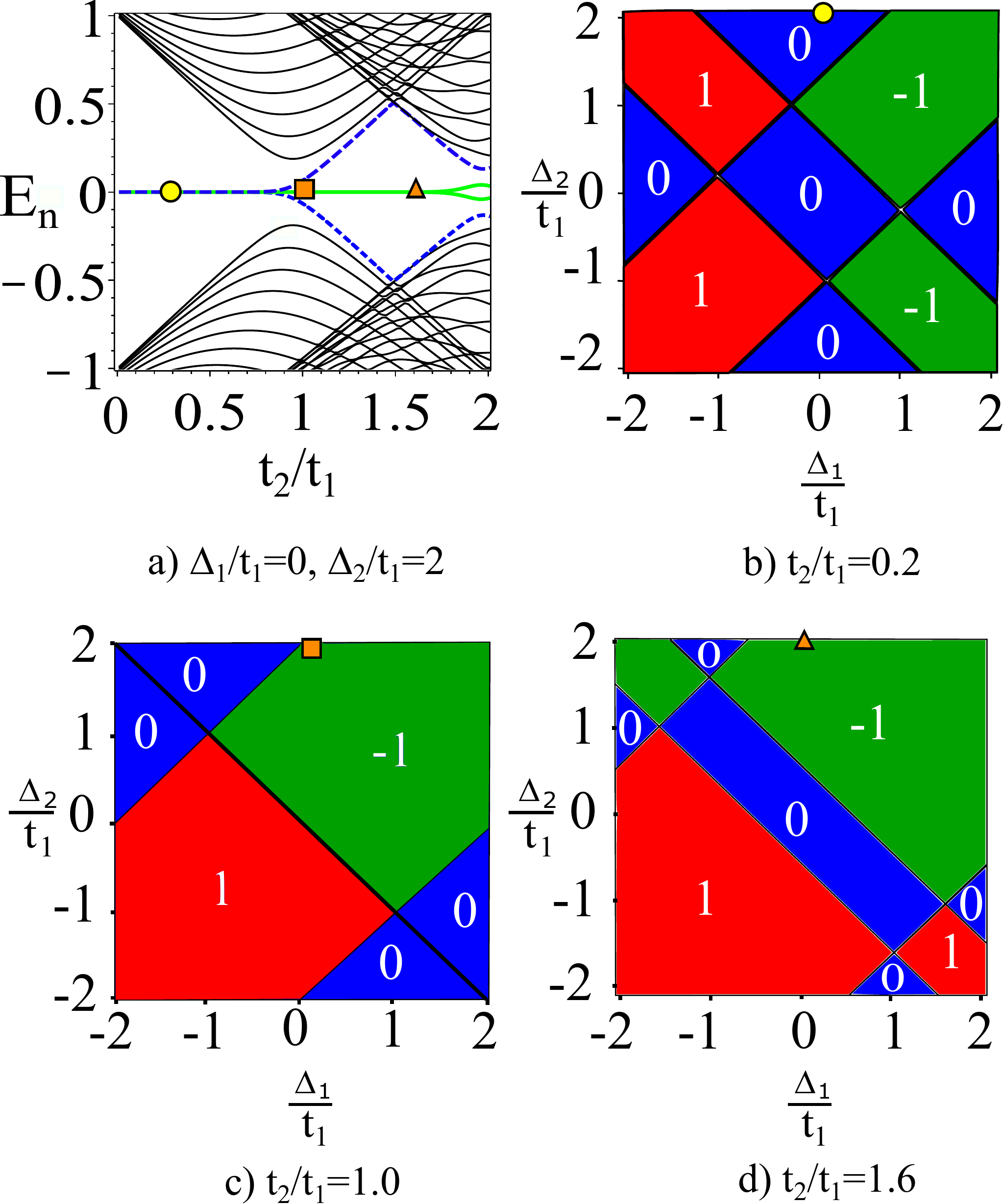}
  \caption{(Color online) Energy spectrum in real space and topological phase diagrams with respect to $W_2$ with $\mu=0$. The green, red and blue regions correspond to $W_2=-1$, $W_2=1$ and $W_2=0$. The numbers in the figures (b), (c) and (d) denotes $W_2$. (a) Energy spectrum as a function of $\frac{t_2}{t_1}$ for the particular point $\frac{\Delta_1}{t_1}=0$ and $\frac{\Delta_2}{t_1}=2$ of the phase diagrams (b), (c) and (d). Here, the black solid lines are the energy of the bulk states while the blue and green lines are the energy of the edges states.  The circle (yellow), square (orange) and triangle (orange) in the energy spectrum (a) indicates the value of the ratio $t_2/t_1$. (b) Topological phase diagram with $\frac{t_2}{t_1}=0.2$. where the circle indicates the point $(0,2)$.  (c)  Topological phase diagram with $\frac{t_2}{t_1}=1$, where the square indicates the point $(0,2)$. (d) Topological phase diagram with $\frac{t_2}{t_1}=1.6$, where the triangle indicates the point $(0,2)$.
   }\label{DiagramaC2muZero}
\end{figure}

\subsection{One half limit,  $ \mu = 0 $}

Since the model in Eq.~\ref{Hibrido} has been classified into the BDI topological class, the $Z$ topological invariant that
should be calculated to characterize the trivial and non-trivial phases is the well known winding number \cite{Maffei}

\begin{equation}\label{winding}
  W_i=\frac{1}{4 \pi i} \int_{0}^{2\pi} dk Tr ( \mathcal{C}_i \mathcal{H}_k^{-1} \partial_k \mathcal{H}_k),
\end{equation}
where $\mathcal{C}_{i=1,2}$ are the symmetry operators and $H_k$ is given by Eq.~\ref{Hk}. The operators $\mathcal{C}_{i=1,2}$ are two matrices that anti-commute with the Hamiltonian ($ \{ \mathcal{C}_i,H_k \}=0 $), and defines two distinct topological indexes $W_{i=1,2}$.

In the case of $\mu=0$, the Hamiltonian, Eq~\ref{Hk}, anti-commutes with two matrices, $\mathcal{C}_1=\sigma_{0} \otimes \sigma_{3}$ and $\mathcal{C}_2=\sigma_{0} \otimes \sigma_{1}$. Here, the matrix $\mathcal{C}_1$ is the chiral operator or sublattice symmetry and $\mathcal{C}_2$  is  the particle-hole symmetry operator. Thus, $W_1$  (or the chiral index $N_1$ in the  Ref~\cite{Wakatsuki}) is related to the sublattice symmetry, and $W_2$ (or the chiral index of Majorana fermion $N_2$ in Ref~\cite{Wakatsuki}) refers to the particle-hole symmetry.

The invariant $W_1$ is related to the Zak's phase $\gamma_n$, such that $\gamma_n=W_1 \pi$. Here, $n$ identifies the index of the occupied band.  The interpretation of the Zak phase and the winding number are physically different. Due to the bulk-boundary correspondence,  the winding number $W_1$ can be interpreted as the number of zero-energy states. The winding number is a quantized number. However, the Zak phase is the Berry's phase picked up by the eigenfunction of Hamiltonian $H_k$ when $k$ is forced to vary by a external perturbation through the entire Brillouin zone. The Zak phase is quantized only when the system possesses inversion symmetry or chiral symmetry \cite{Zak},\cite{Fradkin}.

Besides, as demonstrated by R. Wakatsuki \emph{et al} Ref~\cite{Wakatsuki}, $W_1$  is equal to the number of zero energy states per end of the chain and $|W_2|$ is equal to the number of Majorana zero-energy bound states. When $\Delta_1=\Delta_2=\mu=0$, the chiral operator $\mathcal{C}_1=\sigma_{0} \otimes \sigma_{3}$  becomes $\sigma_{3}$.   This reduction, $\mathcal{C}_1=\sigma_{3}$ is necessary to obtain a quantum phase transition from the hybrid chain to the SSH limit.

%DISCUSSÃO DAS FIGURAS 2 E 3
We calculated $ W_1=\Theta(|\Delta_2+t_2|-|\Delta_1-t_1|+\Theta(|\Delta_2-t_2|-|\Delta_1+t_1|) $ for $\mu=0 $ (see Appendix \ref{SecW}) and the phase diagram results are shown in Fig.~\ref{DiagramaMuzero}. All parameters are expressed in units of $ t_1 $ in the plot.
The results for $W_2 $, can be seen in Fig.~\ref{DiagramaC2muZero}. The analytical expressions for the winding numbers when $ \mu = 0 $ are given in Appendix~\ref{SecW}. As we can see, in Fig~\ref{DiagramaMuzero} and Fig.~\ref{DiagramaC2muZero}, these phase diagrams are presented as function of the parameter $\frac{\Delta_1}{t_1}$ and $\frac{\Delta_2}{t_1}$ for fixed values of $\frac{t_2}{t_1}$.

Fig~\ref{DiagramaMuzero} (a) shows the phase diagram of the pure SSH model without the superconducting correlation terms. In this case, a topological non-trivial phase arises when $|t_2|>|t_1|$ since $W_1=1$ (red region), and a trivial topological phase arises when $|t_1|>|t_2|$ ($W_1=0$) (blue region).

We then proceed our analysis introducing the superconducting correlations $ \Delta_1$ and $ \Delta_2 $ to the SSH model.
Starting from the three different ground states: $ \frac{t_2}{t_1} < 1 $ (topological trivial), $ \frac{t_1}{t_2} = 1 $ (phase transition line) and $ \frac{t_2}{t_1} > 1$ (topological non-trivial phase) in the absence of superconducting correlations ($ \Delta_1 = \Delta_2 = 0 $), we variate the superconducting correlations for each case and the results are presented in Figs.~\ref{DiagramaMuzero} (b), (c) and (d).  The ratio $t_2/t_1$ has been held fixed in each panel.

Fig.~\ref{DiagramaMuzero} (b) shows the case $ \frac{t_2}{t_1} < 1 $  and we observe that the superconducting correlations induce three different topological regions: blue ($ W = 0 $), red ($ W = 1 $), and an extra white region ($ W = 2 $). Moreover, Fig.~\ref{DiagramaMuzero} (c) shows the results along the transition line, $ \frac{t_1}{t_2} = 1 $, in the presence of the parameters $ \frac{\Delta_1}{t_1} , \frac{\Delta_2}{t_1} \neq 0 $, and we get the same three different topological phases as previously obtained, indicated by the same color scheme as in Fig.~\ref{DiagramaMuzero} (b),
however, notice that now, there is a black point located at the origin of the phase diagram. This point represent a gapless point of the SSH model, $ \frac{\Delta_1}{t_1} = \frac{\Delta_2}{t_1} = 0 $, which is consistent with the phase diagram in Fig.~\ref{DiagramaMuzero} (a) at point $t_1=t_2=1$.

%{\color{red}  Mathematically, the invariant $W_1$ is equal to 2 when  $\Delta_i/t_1=0$. This occurs because the presence of particle-hole symmetry in system induces two copies of the SSH model. However, physically, the point $\Delta_i/t_1=0$ represents a quantum phase transition from the superconducting to normal metal phase, where the superconducting gap $\Delta_i$ and the energy spectrum gap goes to zero.  In this sense, we claim that the correct physical results for $\Delta_i/t_1=0$ need to be found with the arguments that,  in the normal phase,  the ground state do not need two copies of the SSH model. Therefore, we do not need the particle-hole subspaces and the invariant $W_1$ is reduced to 1, while the invariant $W_2$ can not be defined any more. }

Finally, the phase diagram shown in Fig.~\ref{DiagramaMuzero} (d) emerges from a topological phase of the SSH model, when $ \frac{t_2}{t_1} = 1.6$.

The results are analogous to those previously obtained, except that, differently from the result obtained by R. Wakatsuki \emph{et al}.~\cite{Wakatsuki},
we have correctly determined the topological phase transition at $ \Delta_1 = \Delta_2 = 0 $, see the red dot at the origin of the phase diagram.

Indeed, the model should fall into the topological phase of the SSH model, $ W_1 = 1 $, when $ \frac{t_2}{t_1} > 1 $ in the absence of superconducting correlations (see the red dot at the origin of the diagram in Fig.~\ref{DiagramaMuzero} (d).

Fig.~\ref{DiagramaC2muZero} (a) shows the energy spectrum in real space for $ N = 50 $ as a function of $ \frac{t_2}{t_1} $ for fixed values $ \frac{\Delta_1}{t_1} = 0 $ and $ \frac{\Delta_2}{t_1}= 2 $.

The zero-energy states located at the ends of the chain are highlighted by green and blue solid lines, and correspond to the orange square and triangle, indicated by the point $\Delta_1=0$ and $\Delta_2=2$ of the phase diagram in Fig.~\ref{DiagramaC2muZero} (c) and (d). On the other hand, the bulk energy states have been highlighted by black solid lines, corresponding to the yellow circle in the energy spectrum of Fig.~\ref{DiagramaC2muZero} (a), are related to the point $\Delta_1=0$ and $\Delta_2=2$ of the phase diagrams (b) and (c) of Fig.~\ref{DiagramaC2muZero}.

The zero-energy edge states solutions are shown in Fig.~\ref{DiagramaC2muZero} (a) and indicated by the red dotted and green solid lines.

The circle indicates the topological non-trivial ground state (topological indexes $W_1 = 2$ and $W_2 = 0$) with four zero-energy states at the ends of the chain, see Fig.~\ref{DiagramaC2muZero} (a). The orange square and triangle shows the topological non-trivial region (topological indexes $W_1 = 1$ and $W_2 =-1$) with two zero-energy states at the ends of the chain, Fig.~\ref{DiagramaC2muZero} (a).

We see that, in the interval $ 0 < \frac{t_2}{t_1} < 1.0 $  of the Fig.~\ref{DiagramaC2muZero} (a) $ W_1=2 $. Therefore,  using the bulk boundary correspondence theorem \cite{Roger}, one realizes that $ W_1=(4)/(2)=2 $ is the number of zero-energy states per end of the chain. Indeed, the white regions in Fig.~\ref{DiagramaMuzero} (b), (c) and (d) indicate two zero energy states per ends of the chain.

Moreover, we see that for $1.0 \leq \frac{t_2}{t_1} \leq 1.8$, there are only two zero-energy states in Fig.~\ref{DiagramaC2muZero} (a) (see the two collapsed green lines in this interval). Again, using the bulk boundary correspondence theorem, we conclude that $ W_1  = 1 $ is the number of zero-energy states per end of the chain
(i.e. $W_1 = (2)/(2)=1$ states/end).

It is know that, given the particle-hole symmetry, a pair of Majorana zero-energy bound state emerge at the ends of the chain when each end of the chain has only one zero-energy state. For instance, on the interval $ 1.0 < \frac{ t_2 }{ t_1 } < 1.8 $, and for $ \frac{\Delta_1}{t_1}=0 $ and $ \frac{\Delta_2}{t_1}=2 $ there is only one Majorana zero-energy bound state per end of the chain, see solid green lines in Fig.~\ref{DiagramaC2muZero} (a). Indeed, the point $ ( \frac{ \Delta_1 }{ t_1 }=0, \frac{ \Delta_2 }{ t_1 } = 2 ) $  from Fig.~\ref{DiagramaC2muZero} (c) and (d) is located within the green region with $ W_2 = -1 $, see orange square and triangle in Figs.~\ref{DiagramaC2muZero} (a), (c) and (d).

Actually, the green and red regions in Fig.~\ref{DiagramaC2muZero} support one isolated Majorana zero-energy bound state  at the edges, while blue regions do not exhibit an isolated Majorana zero-energy bound state. The boundaries of the phase diagrams Fig.~\ref{DiagramaMuzero}  and  Fig.~\ref{DiagramaC2muZero} indicate the phase transitions between topological phases with different number of zero-energy edge states, such that green and red regions exhibit ground states with Majorana zero-energy bound states at the edges of the chain.

%COMPARANDO O NOSSO CASO COM O NAGAOSA
In order to investigate the topological phase transition from the hybrid phase to a pure SSH ground state, we introduced the following parameterizations; $t_1 = -t(1+\eta_1)$, $t_2 = -t(1-\eta_1)$, $\Delta_1= \Delta( 1+\eta_2 )$ and $\Delta_2= \Delta( 1-\eta_2 )$. The case,  $\eta_1=\eta_2$ was studied in Ref~\cite{Wakatsuki}.

We notice that in Ref~\cite{Wakatsuki} the authors studied only the case $\alpha=1$, i.e $\eta_2=\eta_1$.
We remark that the choice $\eta_1=\eta_2$ creates a specific constraint between the hopping and superconducting parameters.

\begin{figure}[t!]
  \centering
  \includegraphics[width=1\columnwidth]{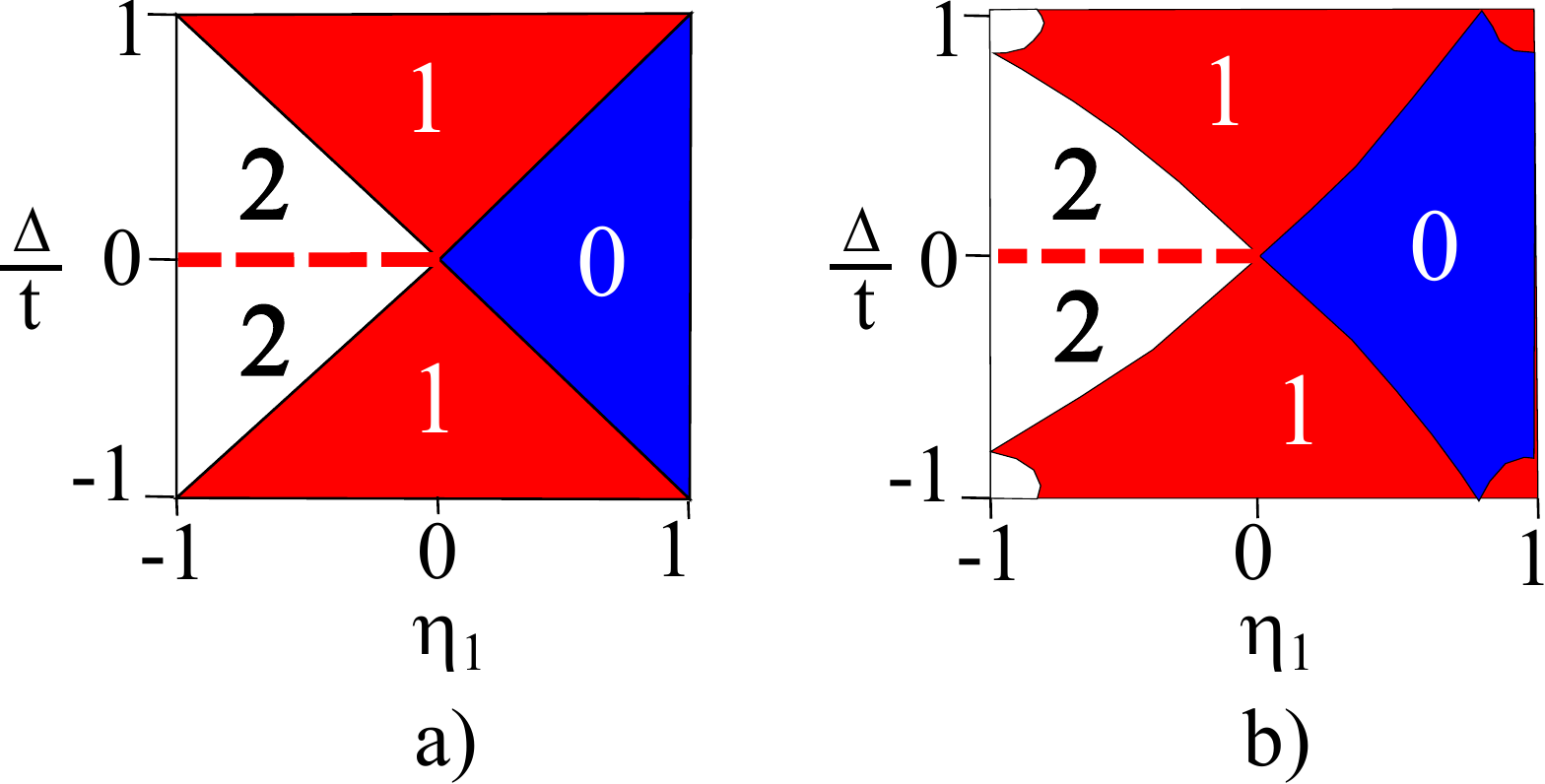}
  \caption{(Color online) Topological phase diagrams with respect to $W_1$ in-plane $\frac{\Delta}{t}-\eta_1$ with $\mu=0$. Parametrization $t_1=-t(1+\eta_1)$, $t_2=-t(1-\eta_1)$, $\Delta_1=\Delta(1+\eta_2)$ and $\Delta_2=\Delta(1-\eta_2)$,\cite{Wakatsuki}. The numbers in the figures denotes $W_1$. Red regions exhibit phase with $W_1=1$, blue regions exhibit a phase with $W_1=0$ and white regions posses $W_1=2$. (a)  Case $\eta_2=\eta_1$. (b) Case $\eta_2=1.5\eta_1$. All points that belong to the traced line at $\frac{\Delta}{t}=0$ and $\eta_1<0$ are into a topologically non-trivial phase of the pure SSH model, since at these points, the winding number  $W^{SSH}=1$. The topological traced line for $\eta_1<0$ separates two topological regions with the same topological invariant $W_1=2$ (white region).}\label{DiagramaMuzeroEtas}
\end{figure}

Replacing the above parametrization in the Hamiltonian \ref{Hibrido}, we obtain a new phase diagram, as seen in Fig.~\ref{DiagramaMuzeroEtas}.
Moreover, Fig.~\ref{DiagramaMuzeroEtas} provides a clear visualization of the topological quantum phase transition from the superconducting hybrid model to SSH model. The dotted line at $\Delta=0$ represents the topological insulating phase of the SSH model with $W^{SSH}=1$, see Fig.~\ref{DiagramaMuzeroEtas} (a) and (b).

In order to compare with previous results, in Fig.~\ref{DiagramaMuzeroEtas} (a) we fixed  $\eta_1=\eta_2$. The phase diagram when $\eta_1 \neq \eta_2$ can be visualized in Fig.~\ref{DiagramaMuzeroEtas} (b). In this case, we fixed $\alpha=1.5$, $\eta_2= \alpha \eta_1$ and $0\leq \eta_1\leq1$.
We can see three different topological phases, red $W_1=1$, blue $W_1=0$, white $W_1=2$ and a red traced line with $W^{SSH}_1=1$ for $\Delta/t =0 $ and $\eta_1<0$. Here $W^{SSH}_1$ is the winding number for the insulating phase of the SSH model.  This result has not been obtained in Fig. 2-(a) of Ref~\cite{Wakatsuki}, where for $\Delta/t=0$ and $\eta<0$, they found $N_1=2$ in their whole SSH-like area.

\subsection{ $ \mu \neq 0 $ }

\begin{figure}[t!]
  \centering
  \includegraphics[width=1\columnwidth]{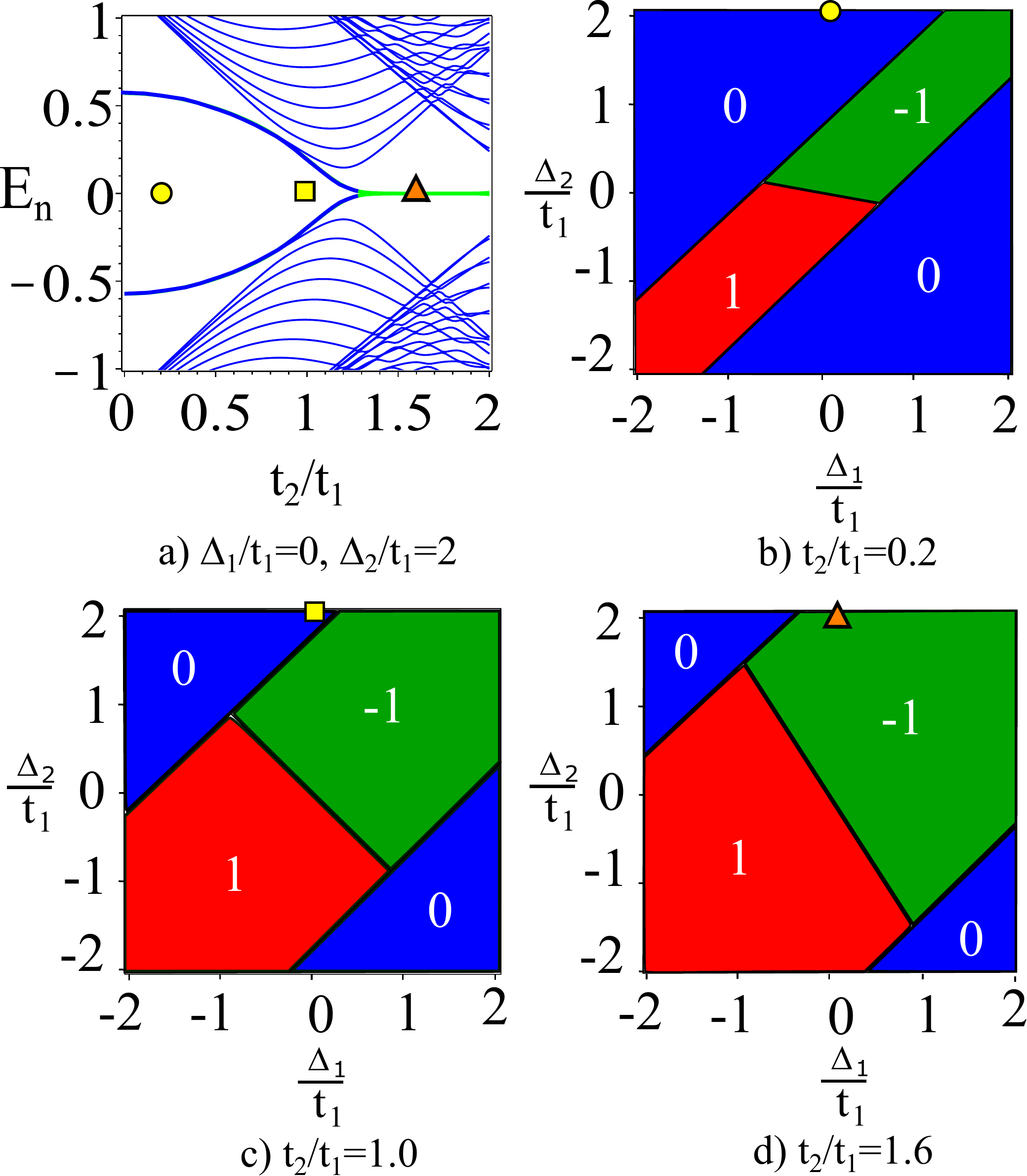}
  \caption{(Color online) Energy spectrum of the finite chain as a function of $\frac{t_2}{t_1}$ and topological phase diagrams with respect to $W_2$ with $\mu=0.9$. The numbers in the figures (b), (c) and (d) denotes $W_2$.
  (a) Energy spectrum as a function $\frac{t_2}{t_1}$ with $\frac{\Delta_1}{t_1} = 0$, $\frac{\Delta_2}{t_1}=2.0$ and $\mu=0.9$. Here, the black solid lines are the energy of the bulk states while the green lines are the energy of the edges states. In this case, we can see zero-energy Majorana bound states only for $1.2<t_2<2.0$ (solid green lines at $E=0$).
  The circle, square and triangle in the energy spectrum (a) indicates the  point $\Delta_1=0$ and $\Delta_2=2$ of the phase diagrams (b), (c) and (d).
  (b) Topological phase diagram with $\frac{t_2}{t_1}=0.2$, (c) Topological phase diagram with $\frac{t_2}{t_1}=1.0$ and (d) Topological phase diagram with $\frac{t_2}{t_1}=1.6$). The point $(0,2)$ of the phase diagrams in (b), (c) and (d) are indicated by  circle (yellow), square (yellow) and triangle (orange).
 Here, green color corresponds to $W_2=-1$, red $W_2=1$ and blue $W_2=0$. }\label{DiagramaMudiferentzero}
\end{figure}

For $| \mu | > 0 $ the sublattice symmetry is explicitly broken and only the particle-hole symmetry ( $\Xi_2=\sigma_{1} \otimes \sigma_{0} \mathcal{K}$) induce the topological index. We computed the number of zero-energy states and $W_2$ (see Eq~\ref{Wmu} in Appendix \ref{SecW}) as a function of $\Delta_1$, $\Delta_2$  and $\frac{\mu}{t_1}$. These results are summarized in figures \ref{DiagramaMudiferentzero}.

Fig.~\ref{DiagramaMudiferentzero} shows the effects of the chemical potential $\frac{\mu}{t_1}$ over the number of zero-energy solutions per end of the chain.

In Fig.~\ref{DiagramaMudiferentzero} (a), we calculated the energy spectrum for a chain with $60$ sites as a function of $\frac{t_2}{t_1}$ for $\frac{\mu}{t_1}=0.9$, $\frac{\Delta_1}{t_1}=0$ and $\frac{\Delta_2}{t_1}=0.4$.

The zero-energy states located at the ends of the chain are highlighted by green solid lines, and correspond to the orange triangle, indicated by the point $\Delta_1=0$ $\Delta_2=2$ of the phase diagram in Fig.~\ref{DiagramaMudiferentzero} (d). On the other hand, the bulk energy states have been highlighted by black solid lines, corresponding to the yellow circle and square in the energy spectrum of Fig.~\ref{DiagramaMudiferentzero} (a), are related to the point $\Delta_1=0$ and $\Delta_2=2$ of the phase diagrams (b) and (c) of Fig.~\ref{DiagramaMudiferentzero}.

Fig.~\ref{DiagramaMudiferentzero} (b) shows the phase diagram in plane $\Delta_1-\Delta_2$ for $\mu/t =0. 9 $ and $\frac{t_2}{t_1}=0.2$.
Again, the regions with $W_2 =0$ are colored with blue color, while the regions with $W_2=-1$ and $W_2=1$ are colored with green and red colors respectively.
The yellow circle indicates the topological trivial ground states ($W_2=0$) for $\mu=0.9$, $t_2/t_1=0.2$, $\Delta_1=0$ and $\Delta_2=2$.

Fig.~\ref{DiagramaMudiferentzero} (c) shows the phase diagram when $\frac{t_2}{t_1}=1$, where the yellow square indicates the topological trivial ground state ($W_2=0$) for $\mu=0.9$, $t_2/t_1=0.2$, $\Delta_1=0$ and $\Delta_2=2$.

Fig.~\ref{DiagramaMudiferentzero} (d) shows the phase diagram for $\frac{t_2}{t_1}=1.6$.  The orange triangle indicates the topological non-trivial ground state ($W_2=0$) for $\mu=0.9$, $t_2/t_1=1.6$, $\Delta_1=0$ and $\Delta_2=2$. We can see clearly that the topological regions with
$W_2 \neq 0 $ increasing when the ratio $t_2/t_1$ increases (i.e.  the size of the red and green regions from the Fig.~\ref{DiagramaMudiferentzero} (b) ($t_2/t_1$ = 0.2), (c)($t_2/t_1$ = 1.0)  (d) ($t_2/t_1$ = 1.6) increases as the ratio ($t_2/t_1$) increase.).

One can compare the  phase diagram for $\mu=0$ (Fig~\ref{DiagramaC2muZero}) with the phase diagram for $\mu=0.9$ (Fig~\ref{DiagramaMudiferentzero}).
The points represented by circle, square and triangles can be used to exemplify the effects of $\mu$.
When we considered $t_2/t_1=0.2$ and $\mu=0.9$, one can see that the circle stays in the trivial phase (blue region in Fig~\ref{DiagramaMudiferentzero} (b) ) and we noted yet a great expansion of the blue region. Now, considering $t_2/t_1=1$ and increasing $\mu$ from zero to 0.9, we observed that the blue region increases too and the square now is in the trivial phase,  see Fig~\ref{DiagramaC2muZero} (c) and  Fig~\ref{DiagramaMudiferentzero} (c). In summary, the chemical potential tenders to increasing the blue region and therefore is prejudicial to the nontrivial topological phase.

\section{Simulation of the zero energy states around the domain wall and edges of the hybrid chain } \label{sec4}

\begin{figure}[t!]
\centering
\includegraphics[width=1\columnwidth]{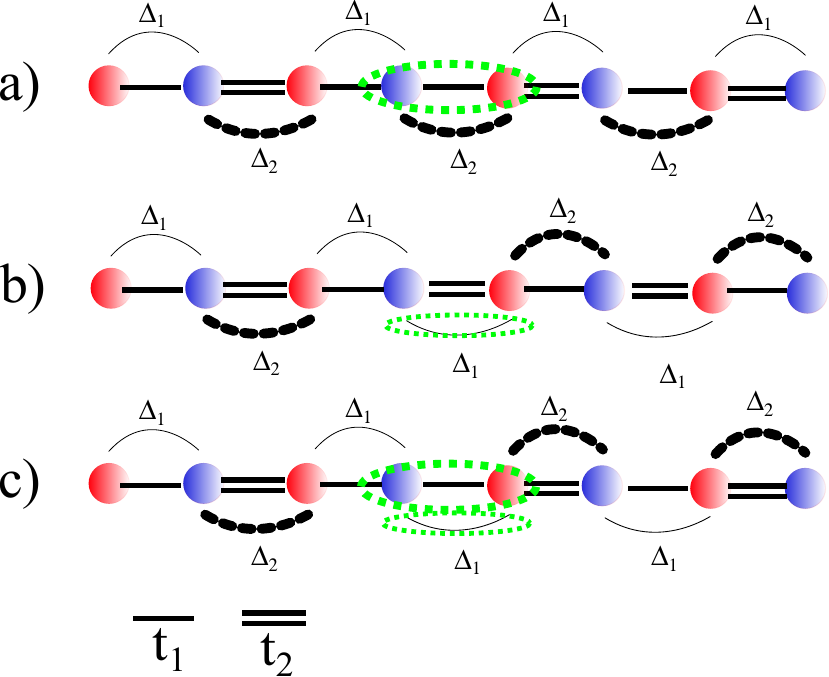}
\caption{(Color online) The hybrid chain in the presence of a kink at the middle of the chain. In panel (a), we considered a kink only in hopping terms, see green dotted circle at the middle of the chain. In panel (b), we considered a kink only in the superconducting correlations, while in panel (c), we considered a kink in both, hopping and superconductor correlation parameters. }\label{kink}
\end{figure}

In this section, we consider a kink at the middle of the hybrid chain \cite{Jin}. In Fig.~\ref{kink} (a), we considered a kink only in the hopping terms, see green dotted circle. In Fig.~\ref{kink}  (b), we considered a kink only in the superconducting correlations, while in Fig.~\ref{kink}  (c), we considered a kink in both, hopping and superconducting pairing parameters.

The Hamiltonian of the hybrid chain in the presence of the kink is given by

\begin{eqnarray}\label{Hamiltonkinks}
  H &=& \psi^{\dagger} h \psi
\end{eqnarray}
where,
\begin{eqnarray}\nonumber
\psi^{\dagger}&=&(c^{\dagger}_{A1},...,c^{\dagger}_{AN};c^{\dagger}_{B1},...,c^{\dagger}_{BN};  c_{A1},...,c_{AN};c_{B1},...,c_{BN} ),
\end{eqnarray}
and

\begin{equation}\label{HKink}
  h=\left(
        \begin{array}{cccc}
          0 & T & 0 & \Delta \\
          T^{\dag} & 0 & -\Delta^{\dag} & 0 \\
          0 & -\Delta & 0 & -T \\
          \Delta^{\dag} & 0 & -T^{\dag} & 0 \\
        \end{array}
      \right).
\end{equation}

The matrices  $T$ and $\Delta$ depends on the kink, see Appendix \ref{Matrices}. For instance, for a kink only in the hopping term, the matrices $T$ and $\Delta$ are given by

\begin{align}
 T_{i,j}
&=
\begin{cases}
\delta_{i,j} t_1 +  \delta_{i-1,j} t_2,
\quad
& i,j \leq N/2
\\
\delta_{i,j} t_2 +  \delta_{i-1,j} t_1,
\quad
&i,j > N/2
 \end{cases}\label{HoppingMatrix}
\end{align}
and

\begin{equation}\label{DeltaKinka}
  \Delta_{i,j}=\delta_{i,j} \Delta_1 +  \delta_{i-1,j} \Delta_2
\end{equation}

\begin{figure*}
  \centering
  \includegraphics[width=13cm]{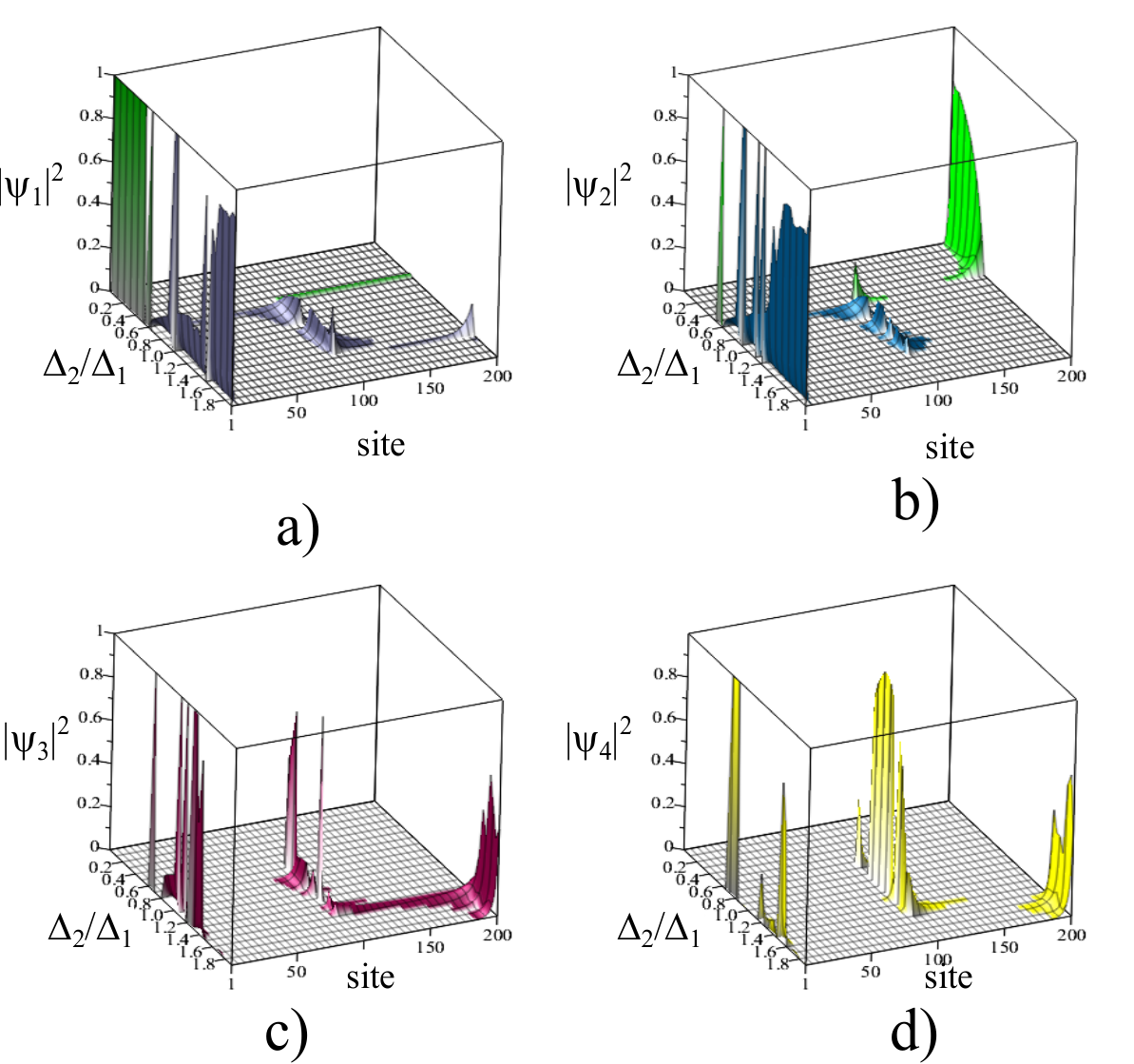}
  \caption{(Color online) Probability density of the  zero-energy states as a function of the ratio
$\Delta_2/\Delta_1$ and the site position. (a)$|\psi\Psi_1|^2$ , (b)$|\psi_2|^2$, (c) $|\psi_3|^2$ and (d) $|\psi_4|^2$ }\label{FourFirstHopp}
\end{figure*}

Otherwise, when we considered a kink in hopping and superconducting pairing terms, the matrix $T$ remains equal to \ref{HoppingMatrix}, while
the matrix $\Delta$ should be replaced by

\begin{align}
 \Delta_{i,j}
&=
\begin{cases}
\delta_{i,j} \Delta_1 +  \delta_{i-1,j} \Delta_2,
\quad
& i,j \leq N/2
\\
\delta_{i,j} \Delta_2 +  \delta_{i-1,j} \Delta_1,
\quad
&i,j > N/2
 \end{cases}
\end{align}

\begin{figure*}
  \centering
  \includegraphics[width=12cm]{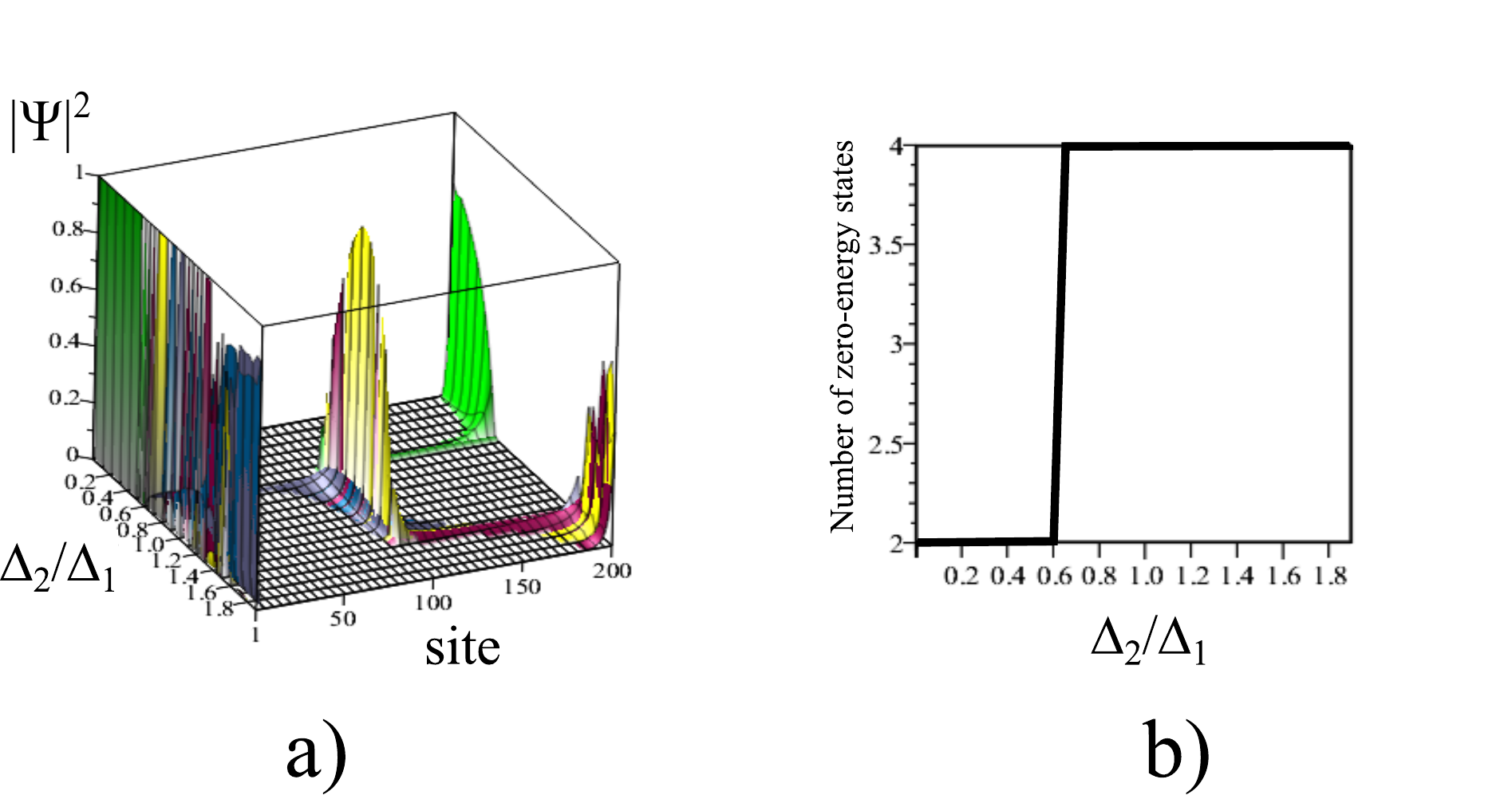}
  \caption{(Color online) Probability density of the  zero-energy states. (a) Probability density of all zero-energy states together in the same picture,
  (b) the number of the zero energy states as a function od the ratio $\Delta_2/\Delta_1$. }\label{ZeroFirstHopp}
\end{figure*}

\begin{figure*}
  \centering
  \includegraphics[width=12cm]{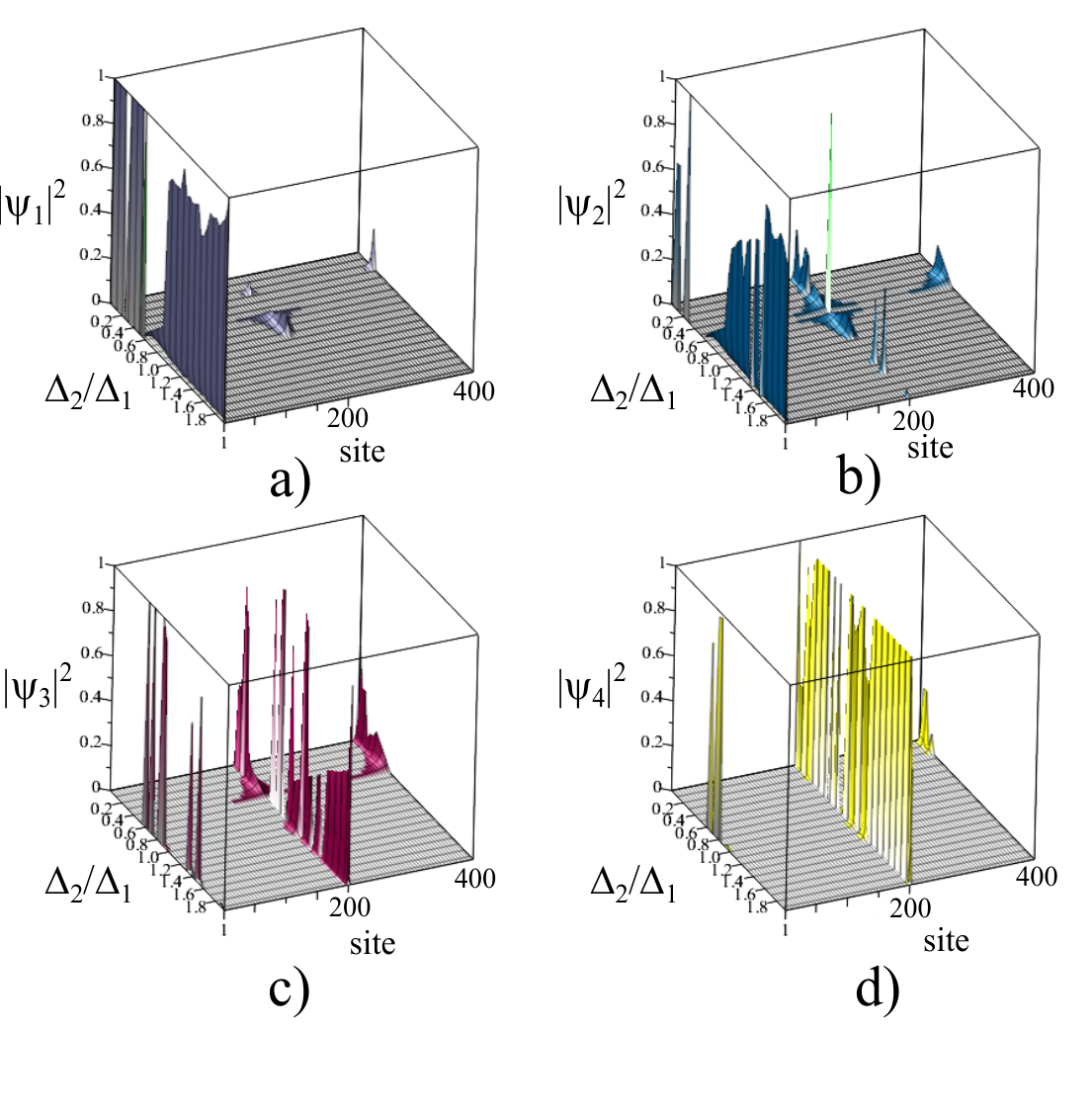}
  \caption{(Color online) Probability density of the Four Zero-Energy states as a function of the ratio
$\Delta_2/\Delta_1$ and the site position. (a)$|\psi_1|^2$ , (b)$|\psi_2|^2$, (c) $|\psi_3|^2$ and d)$|\psi_4|^2$ }\label{FourFirstHoppDelta}
\end{figure*}

\begin{figure*}
  \centering
  \includegraphics[width=12cm]{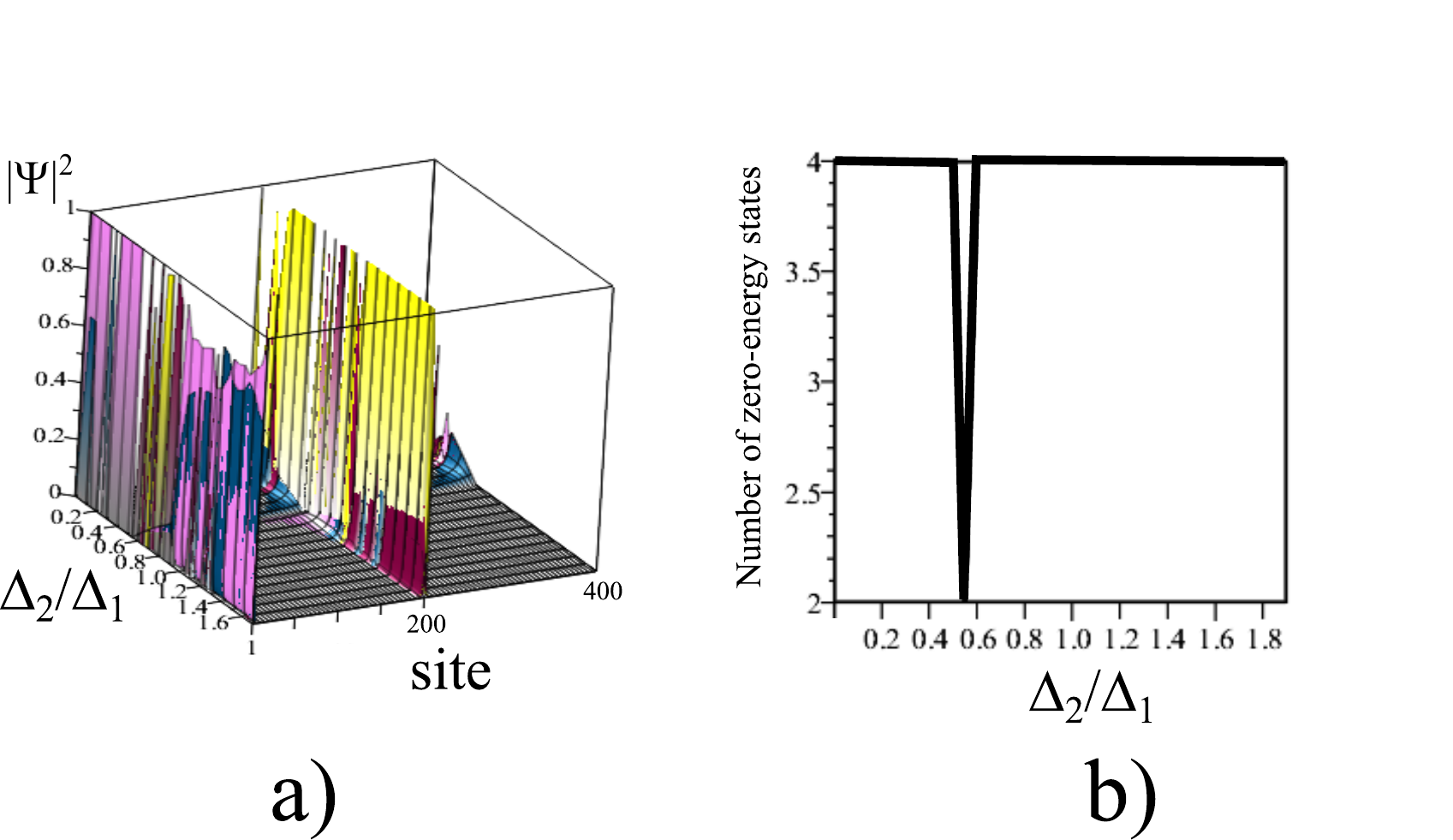}
  \caption{(Color online) Probability density of the zero-energy states of the chain in presence of the kink. (a) Probability density of all zero-energy states together in the same picture, (b) the number of the zero energy states as a function od the ratio $\Delta_2/\Delta_1$. }\label{ZeroFirstHoppDelta}
\end{figure*}

\begin{figure*}
  \centering
  \includegraphics[width=12cm]{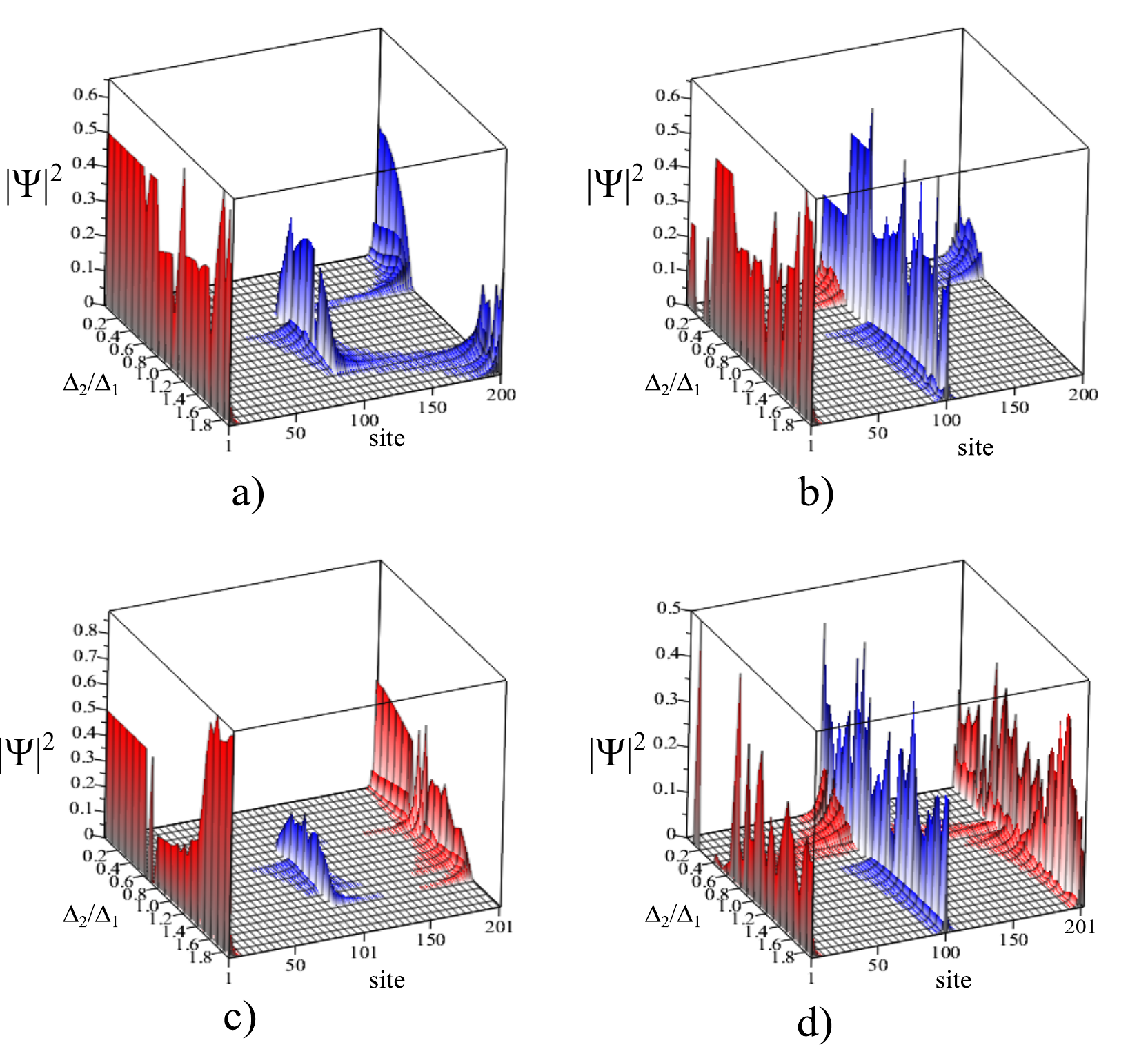}
  \caption{(Color online) Probability density of the zero-energy states as a function of the ratio
$\Delta_2/\Delta_1$ and the site position. Blue and Red color denotes the subreds A nd B, respectively. (a) Chain with $200$ sites with kinks only in the hopping terms. (b) Chain with $200$ sites and kink in hopping and pairing terms. (c)  Chain with one additional site at the right edge of the chain and with kink only in the hopping terms , d) Chain with additional site at the right edge of the chain and with kink in hopping and pairing terms. }\label{Subllatices}
\end{figure*}

\begin{figure}
  \centering
  \includegraphics[width=1\columnwidth]{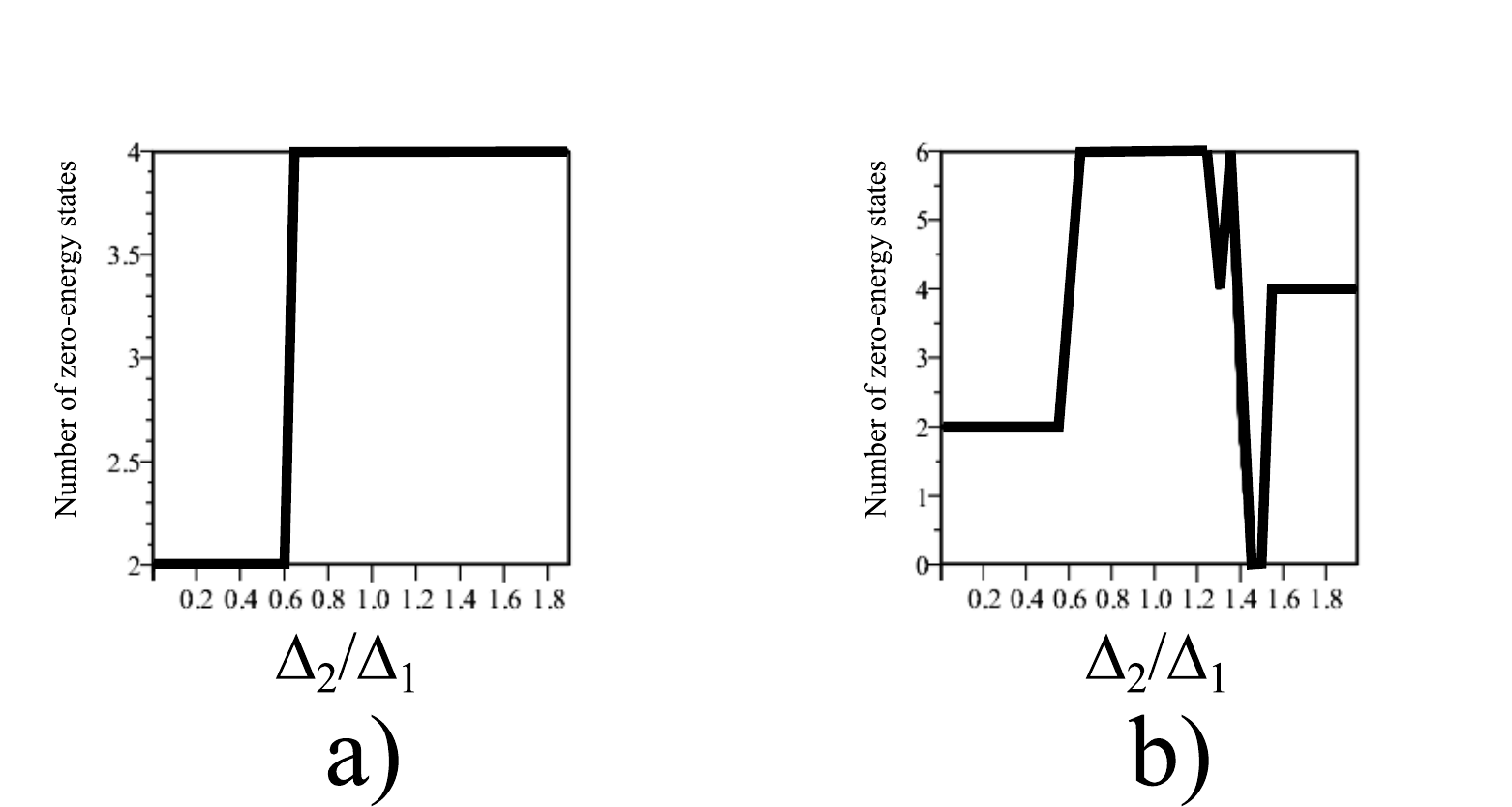}
  \caption{(Color online) Number of zero-energy states as a function of the parameter $\Delta_2/\Delta_1$. a) kink in hopping terms with $N=200$. b) Kink in hopping terms with a additional site $N =201$.} \label{numeros1add}
\end{figure}

We simulated the situations depicted in Fig.~\ref{kink} (a), (b) and (c). In all simulations, we calculated the probability density of the zero-energy states as a function of the site position and the parameter $\Delta_2/\Delta_1$.
The results of these simulations are shown in Figures \ref{FourFirstHopp}-  \ref{numeros1add}.

 Physically, when the space symmetry is broken, for instance,  at the kink, zero-energy bound states can emerge around it \cite{Wakatsuki}.
The emergence of zero-energy solutions around the middle of the chain occurs due to the domain wall, which separates two sectors of the chain with different topological index.
As we mentioned above, in this section, we studied two types of the kink at the middle of the chain; The first case takes into account one kink only in the hopping terms. The second case, considers the kink in both, hopping and superconducting pairing terns. We neglected the case with kink only in superconducting order parameters (fig~\ref{kink} (b)) because we observed that this situation does not induce zero-energy bound states around the middle of the chain.

{\it Kink in the hopping :---} In the first case, we considered the chain with kink only in the hopping terms and for $t_2/t_1=1.5$.
In the last part of this section, we will generalize the results introducing two dimerizations parameters $\eta_1$ and $\eta_2$, as defined in \cite{Wakatsuki}.

The Hamiltonian of the system in the presence of the kink only in the hopping term, and under open boundary conditions, possesses zero-energy solutions, see fig~\ref{FourFirstHopp} and \ref{ZeroFirstHopp}.  The eigenvectors related to these zero-energy solutions can be used to obtain the probability density $|\psi_j(i)|^2=u_i^*u_i$, where $u_i$ is the ith component of the eigenvalue.

As we can see in fig~\ref{FourFirstHopp}, in the presence of the kink, the chain with open boundary conditions can exhibit two or four zero-energy solutions, named as $\psi_1$, $\psi_2$, $\psi_3$, $\psi_4$. We notice yet that the number and location of these zero-energy solutions depend on the ratio $\Delta_2/\Delta_1$, see  \ref{ZeroFirstHopp}.   For instance, for $0<\Delta_2/\Delta_1<0.6$, the chain exhibits two zero-energy states, one around the left edge and other around the right edge, see the green regions in fig \ref{FourFirstHopp} (a) and (b). These solutions are majority distributed around the edges and the middle of the chain, as we can see in figs \ref{FourFirstHopp} (a), (b), (c) and (d).

Besides the number of zero-energy states, fig.~\ref{FourFirstHopp} shows the localization of these four zero-energy solutions as a function of the ratio $\Delta_2 / \Delta_1$ and site position of the chain. Now, we will discuss the localization of these zero-energy states as a function of the ratio $\Delta_2/\Delta_1$.
For $ 0 < \Delta_2/\Delta_1 < 0.6 $, the solutions $\psi_1$ and $\psi_2$ are majority localized around the left and right edge of the chain, respectively, see fig.~\ref{FourFirstHopp} (a)-(b). On the other hand, when $ 0.6 < \Delta_2/\Delta_1 < 1.5 $, the solutions $\psi_1$ and $\psi_2$ appears around the left edge and around the middle(kink) of the chain, see fig.~\ref{FourFirstHopp} (a)-(b). Finally, for $\Delta_2/\Delta_1 > 1.5$, $\psi_1$ and $\psi_2$ emerges only around the left and right edges of the chain.

On the interval $0.6<\Delta_2/\Delta_1 < 1.5$, the zero-energy states $\psi_3$ and $\psi_4$ are localized around the  left edge and middle of the chain,  while for $\Delta_2/\Delta_1 >1.5$, we can see both solutions localized at the right edge, see Figs~\ref{FourFirstHopp} (c)-(d).

We can observe the degree of degeneracy of the zero-energy states in Fig~\ref{ZeroFirstHopp}. Fig~\ref{ZeroFirstHopp} (b) shows the number of these zero-energy states as a function of the ratio $\Delta_2/\Delta_1$ and site position. Here, we considered $t_2/t_1=1.5$.

At this point, it is important to comment that these zero-energy modes exhibits the following property, at the critical points (e.g. $\Delta_2/\Delta_1=0.6$) the zero-energy solutions tend to spread in the bulk.  The value of $\Delta_2/\Delta_1=0.6$, where the zero-energy modes penetrates in the bulk, corresponds (or are very close) to the critical point of the system, in which the gap energy goes to zero. This critical point separates two distinct topological regions with different numbers of zero-energy states, as we can see in Fig~\ref{ZeroFirstHopp} (a). We point out that, close to the critical point, the zero energy state possesses a penetration depth that can be associated with the correlation length $\xi $ \cite{Rufo}.
The correlation length $\xi$ diverges at the critical point, following the law $\xi^{-\nu} \propto (\lambda-\lambda_c)^{-\nu}$, where $\lambda = \Delta_2 / \Delta_1$. Here $\nu =1$ is the spatial correlation critical exponent. This behavior, at the critical point,  is independent of the number of sites.

Following the discussion, the solutions $\psi_3$ and $\psi_4$ are localized around the left edge and middle on the interval $ 0.6 < \Delta_2/\Delta_1 < 1.5 $, and only at right edge for $\Delta_2/\Delta_1 > 1.5 $, see fig.~\ref{FourFirstHopp} (c) and (d). Note that, these solutions do not appear on the interval $ 0.0 < \Delta_2/\Delta_1 < 0.6$. This particular behavior, of the zero-energy solutions $\psi_3$ and $\psi_4$, guarantees that only one zero-energy solution appears around the left($\psi_1$) and right($\psi_2$) edges of the chain.

Here, we observed another interesting characteristic of our anisotropic model with open boundary conditions. On the interval $0.6<\Delta_2/\Delta_1<1.5$, the zero-energy states do not appear around the right edge, as we can see in fig.~\ref{FourFirstHopp} (a)-(d). This behavior appears only in the chain subject to open boundary conditions, as we will discuss in the section below. This result occurs because on the interval $0.6<\Delta_2/\Delta_1<1.5$, the second half of the chain ( right side of the chain relatively to kink) is in a trivial phase. We remark that, in the presence of a kink, the systems behave like two chains that have been glued together exactly at the kink. Therefore, when the second part is in a trivial topological phase, we cant observe zero-energy state around the right end of the chain.

Now, we will clarify the relationship between the number/localization of these zero-energy states and the emergence of one isolated Majorana zero-energy bound state or two pairs of zero-energy states.

In order to clarify the nature of the zero-energy modes, we plotted all zero-energy solutions together, as we can see in Fig.~\ref{ZeroFirstHopp} (a). In Fig.~\ref{ZeroFirstHopp} (a), there are two zero-energy fermionic states when $\Delta_2/\Delta_1 < 0.5$  (green regions). In this case, there is one zero-energy state around the left and other zero energy-state around the right side of the chain. The first state comes from  $\psi_1$ and the other comes from $\psi_2$. We call attention to the fact that, for a finite dimerized Kitaev chain with $N$ sites, the zero-energy states $\psi_1$ and $\psi_2$ overlap and are no longer eigenstates, but hybridize into two eigenstates $\Psi_1=(\psi_1 + i\psi_2)/2$ and $\Psi_2=(\psi_1 - i\psi_2)/2$. The energy of $\Psi_1$ and $\Psi_2$ oscillates around the zero as a function of the $N$. The overlapping between the $\psi_1$ and  $\psi_2$, which in fact will destroy the nonlocal character  of these states, goes to zero when $(L=N)>>\xi$, where $\xi$ is the penetration depth of these states \cite{Dominguez} .

Therefore, when $L>>\xi$ the solutions $\psi_1$ and $\psi_2$ are localized around the left and  right edge and can be interpreted as a Majorana zero-energy bound state. The emergence  of a Majorana zero-energy bound state depends on the many-body ground state degeneracy of these zero-energy modes. Note that, one cannot build a fermionic Fock space out of an odd number of Majorana modes, because they are linear combinations of particles and holes. Rather, we can define a single fermionic operator out of both Majorana end modes at the left and right edges of the chain. Therefore, the Hilbert space we can build out is hence inherently nonlocal \cite{Bernevig} .

In the regions where the values of ratio $\Delta_2/\Delta_1$ induces four zero energy states, see blue, pink and yellow regions in Figs \ref{FourFirstHopp} and \ref{ZeroFirstHopp} ), there is a degeneracy of degree four.  For instance, for $0.5<\Delta_2/\Delta_1<1.5$, there are four zero-energy states around the left edge and the middle of the chain, while for $\Delta_2/\Delta_1>1.5$ these zero-energy states are localized only around the right edge.
Four zero-energy states combine to form an unconventional fermion mode that in general can not be used to build q-bits to store information to realize robust quantum computation\cite{Bernevig}.

\textbf{ It is very interesting that by a fine-tuning of the superconducting order parameter, one can be able to transit between three types of fermionic fractionalization on the chain; 1)  One zero-energy states around the left edge and other around the right edge, 2) two zero-energy state around left edge and two zero-energy around middle or  3) two zero-energy stats around the left and more two around the right edge. This was only possible because we did not  consider  constraints between pairing  and  hopping terms, as taken by \cite{Wakatsuki}.}

{\it Kink in the hopping and pairing terms :---}When we allow a kink in the hopping and superconducting pairing terms, see Fig~\ref{kink} (c), the chain passes to exhibit four zero-energy states, as we can see in Fig~\ref{ZeroFirstHoppDelta} (b).
On the interval $0<\Delta_2/\Delta_1<0.5$ the chain possess two zero-energy states at the left edge, four zero energy states at the middle and two zero-energy states at the right edge, as we can see in Fig~\ref{FourFirstHoppDelta} (a)-(d). Now, for $\Delta_2/\Delta_1>0.5$,  we found two zero-energy states around the left edge of the chain and four around the middle of the chain, see Fig~\ref{FourFirstHoppDelta} (a)-(d).
In this case, the right edge do not exhibits zero-energy states. On interval $0<\Delta_2/\Delta_1<2$, we can not identify only one pair of Majorana zero-energy bound states, since in all cases, at least two zero-energy states occupy the edges and the middle of the chain. The kink in pairing terms increases the degeneracy for all values of $\Delta_2/\Delta_1$, therefore, in this case, the chain exhibits at least two pair of Majorana zero-energy bound states, see Fig~\ref{FourFirstHoppDelta} (a)-(d).
The general zero-energy state is a linear combination of the solutions $\psi_1$, $\psi_2$, $\psi_3$ and $\psi_4$. Again, for a $\Delta_2/\Delta_1>0.5$, there is no zero-energy solution around the right edge of the chain.

{\it Sublattice characteristics of the zero-energy states:---} The zero-energy states in Figs~\ref{FourFirstHopp} and \ref{FourFirstHoppDelta} are localized around the left, right or middle of the chain. However, in which sublattice these states are localized? In fig~\ref{Subllatices}, we show the probability density of these zero-energy states, where the red region denotes the states localized in the sublattice A, while the blue color indicates the states localized around the sublattice B. One can see that, in the case of a kink only in the hopping terms, the states around the left edge of the chain have been colored by red color because these states are localized only in the sublattice A. However, the zero-energy states around the middle and right edge of the chain are colored by blue color, since these states are localized only in sublattice B.  On the other hand, when we considered kink in pairing terms, see fig \ref{Subllatices} (b), we notice that on the interval $0<\Delta_2/\Delta_1<0.4$,  the chain exhibits zero-energy in the sublattice B around the sites  $99$, $101$ and on the right edge of the chain.

{\it Additional site at the right side of the chain:---}We show the zero energy states of the chain in the presence of one additional site (one additional sublattice $A$ at the right side of the chain),  in Fig~\ref{Subllatices}. Now, the chain has $201$ sites. We connected this additional site trough the hopping $t_1$ and pairing $\Delta_2$. Comparing fig~\ref{Subllatices}(a)-(b)(without the additional site) and Fig~\ref{Subllatices} (c)-(d) (with the additional site), one can see now in Fig~\ref{Subllatices} (c)-(d)  two additional zero-energy states around the right edge of the chain. The additional site changes the number of zero energy-states from four to six, as we can see in Fig~\ref{numeros1add} (b). It is interesting to see that one additional site, at the right edge, changes the character of the zero-energy states around this end of the chain. Without the additional site, these states are located only around the sublattice B, however, after adding one site in the right edge,  the zero-energy states around the right edge changes to sublattice A. In this case, both edges passes to exhibits zero-energy states in same sublattice A (red regions), see Fig~\ref{Subllatices} (c)-(d). Note that, the region $0<\Delta_2/\Delta_1<0.6$ possesses two zero energy states exactly equal to the chain without the additional site, see Fig~\ref{numeros1add} (a). In this region, we have a Majorana zero-energy state, exactly equal as we found in previous results. The main difference between the chains with 200 and 201 appears on the interval $0.6<\Delta_2/\Delta_1<1.5$. In this interval,  for a chain with 201 sites, we can find six zero-energy states, where two of these additional states are localized around the right edge, see Figs~\ref{Subllatices} and \ref{numeros1add} (b). These states do not exist for the chain with 200 sites.

{\it Role of boundary conditions:---}All calculations above have been done using open boundary conditions.
We point out that, when we considered the periodic boundary conditions, we observed two additional zero-energy states in the energy spectrum. It occurs because the periodic boundary conditions are implemented by connecting the last and first sites of the chain trough the hopping and pairing terms. When we try to connect the edges of the chain, a second kink is generated. For objectivity, we do not show these results.

%{\it Phase diagram: Number of zero-energy states :---} In fig we show the number zero-energy states in plane $\Delta_2/\Delta_1$-$t_2/t_1$. It is easy to see that the path $\Delta_2/\Delta_1 = t_2/t_1$, in fig~\ref{?}, correspond to the constraint $\eta_1=\eta_2$ in relation the dimerization parameters that was  considered in \cite{Wakatsuki}.

\section{Conclusion}\label{sec5}

In this work, we studied the fermionic fractionalization that emerges in the anisotropic superconducting Su-Schieriffer-Heeger (SSH) model.

The hybrid SSH model exhibits  two distinct discrete symmetries for zero chemical potential and therefore, in this case, these two symmetries (chiral and particle-hole) allow to calculate two distinct topological invariants $W_1$ and $W_2$, where the first can be associated to the number of the zero-edge states per end of the chain and the last tell us if a Majorana zero-energy bound state reside or not at the end of the chain. These results have been confirmed by the calculation of the number of the zero-energy edge states at each case of the phase diagrams obtained trough these two topological invariants.
We also studied the effects of a finite chemical potential over phase diagrams and the existence of fermionic fractionalization, like Majorana zero-energy bound states. Differently from a previous works \cite{Wakatsuki}, the phase diagrams of our hybrid model was correctly reduced to the limit of a pure SSH ground state.

We found that a topological phase transition from a topological non-trivial phase of the hybrid chain to a non-trivial SSH topological phase can be induced for the limit $\Delta_1 \rightarrow 0$, $\Delta_2  \rightarrow 0$, $\mu \rightarrow 0 $ and $t_2/t_1>1$.

In the final part of this work, we simulated the behavior of the zero-energy states around the edges and the domain wall. After creating a domain wall through a kink, we have diagonalized the Hamiltonian in the real space for $200$ sites. We obtained the zero energy solutions around the edges and the kink. We observed that the superconducting correlations dictate the existence of these zero energy states around the domain wall, such that, for some specific values of these correlations, the zero energy states disappear from the middle and become majority localized  around the ends of the chain.

\begin{acknowledgments}
H. C. would like to thank T. Domanski for a helpful conversation. The authors wish to thank CNPQ and FAPEMIG for partial financial support. Griffith M. A. R. would like to thank Capes for postdoctoral fellowship.
\end{acknowledgments}

%%%%%%%%%%%%%%%%%%%%%%%%
% Nota do Lizardo:

\appendix

\section{Analytical derivation of the winding numbers}
\label{SecW}
In this section we provide the analytical expressions for the winding numbers.
By definition, they are calculated from Eq.~\ref{winding},
where $ \mathcal{ H }_k $ is given by Eq.~\ref{Hibrido2}
and it is explicitly expressed here in its matricial form,
\begin{eqnarray}
  \mathcal{ H }( k )
  =
  \begin{pmatrix}
    -\mu & z & 0 & w \\
    z^{ * } & - \mu & -w^{ * } & 0 \\
    0 & -w & \mu & -z \\
    w^{ * } & 0 & -z^{ * } & \mu
  \end{pmatrix}
  \, ,
\label{EqHk}
\end{eqnarray}
with the parameters $ z $ and $ w $
given in terms of
the hoppings strengths and the superconducting gaps,
\begin{eqnarray}
z( k )
& = &
t_1 + t_2 e^{ - i ka }
\, \\
w( k )
& = &
-\Delta_1 + \Delta_2 e^{ - i ka }
\, .
\label{Eqz&w}
\end{eqnarray}
Notice that presently we take
$ a \neq 1 $ for the sake of clarity.

For the particular case of $ \mu = 0 $,
the winding number can be expressed as

\begin{eqnarray}
\label{winding}
  \mathcal{W}
  & = &
  \mathcal{W}_1 + \mathcal{W}_2
  \nonumber \\
  & = &
  \sum_{ i = 1, 2}
  \mbox{ Tr }
  \int_{0}^{2 \pi a }
  \frac{dk }{ 4 \pi i } \,
  \mathcal{C}_i \mathcal{ H }_k^{-1} \partial_k \mathcal{ H }_k
  \nonumber \\
  & = &
  -
  \sum_{ i = 1, 2 }
  \int_{ 0 }^{ 2 \pi a }
  \frac{ dk }{ 2 \pi i } \,
  \partial_k \log \det z_i
  \, ,
  \label{EqW}
\end{eqnarray}
where

\begin{eqnarray}
  z_1
  & = &
  A_1
  + B_1 e^{ - i k a }
  \, \\
  z_2
  & = &
  A_2
  + B_2 e^{ - i k a }
\, .
\label{Eqz1z2}
\end{eqnarray}
with $A_1=t_1-\Delta_1$,  $B_1=t_2+\Delta_2$,  $A_2=-t_1-\Delta_1$ and  $B_1=\Delta_2 - t_2$

Now, the integrals in Eq.~\ref{EqW} can be easily calculated making use of
\begin{equation}
  \int_{ \frac{ - \pi }{ a } }^{ \frac{\pi }{ a } }
    dk
    \frac{ B_i e^{ - i k a }
          }{
              A_i + B_i e^{ - i k a }
            }
    =
    \frac{ 2 \pi }{ a }
  \, ,
  \mbox{ if $ \left|
  % \frac{ B }{ A }
  B_i / A_i
  \right| > 1 $  }
  \, .
\end{equation}
Pluging this result in Eq.~\ref{EqW}, we get
\begin{eqnarray}
  \mathcal{ W }
  &=&
  \Theta\left(|\Delta_2 + t_2| - |  t_1 -\Delta_1 |  \right) \\ \nonumber
  &+&
  \Theta\left( |\Delta_2 - t_2 |- | - t_1 - \Delta_1| \right)
  \, ,
\end{eqnarray}
where $ \Theta( x ) $ denotes the Heaviside step function.

Moreover, making use of the mappings
$ t_1 = - t \left( 1 + \eta_1 \right)  $
and
$ t_2 = - t \left( 1 - \eta_1 \right) $,
where $ t $ denotes their mean value
and
$ \eta_1 $
is the absolute difference between $ t_1 $ and $ t_2 $
divided by $ t $,
and also that
$ \Delta_1 = - \Delta \left( 1 + \eta_2 \right)  $
and
$ \Delta_2 = - \Delta \left( 1 - \eta_2 \right)  $,
in a similar fashion,
one can show that the winding number reduces to
\begin{equation}
 \mathcal{ W }
  =
  \Theta\left( \Delta - t \eta_1 \right)
  +
  \Theta\left( - \Delta - t \eta_1 \right)
  \, .
\end{equation}
In this particular case,
notice that the results
does not depend on the difference between
$ \Delta_1 $ and $ \Delta_2 $.
Moreover, for the even more strict case
when $ \eta_1 =  \eta_2 $,
this result is identical
to the one obtained
by R. Wakatsuki \emph{et al}.~Ref~\cite{Wakatsuki}
and the phase diagram analisys
presented in this paper constitutes
a generalization of their previous investigation.

For the case $ \mu \neq 0 $,
the topological number
is given by Ref~\cite{Wakatsuki},
\begin{equation}
\mathcal{ W }
  =
  - \int_{ \frac{ \pi }{ a } }^{ \frac{ \pi }{ a }  }
  \frac{ dk }{ 2 \pi i }
  \,
  \partial_k  \log Z( k )
  \, ,
\end{equation}\label{Wmu}
where
\begin{equation}\label{Zk}
  Z(k) = \mu^2+(z(k)-w(k))(z(k)^*+w(k)^*).
\end{equation}

\section{Matrices $T$ and $\Delta$}
\label{Matrices}

The matrices $T$ and $\Delta$ represent the hopping and superconducting connections between different sublattices in the real space. These matrices are $N \times N$ matrices, where $N=2n$ is the number of sites and $n$ is the number of unit cells. For a chain with $N=8$ sites the matrices $T$ and $\Delta$ possess the following form,
\newline
\newline
\begin{equation}\label{HKink}
  T=\left(
        \begin{array}{cccc cccc}
          t_1 & 0 & 0 & 0 & 0 & 0 & 0 & 0\\
          t_2 & t_1 & 0 & 0 & 0 & 0 & 0 & 0 \\
          0 & t_2 & t_1 & 0 & 0 & 0 & 0 & 0 \\
          0 & 0 & t_2 & t_1 & 0 & 0 & 0 & 0 \\
          0 & 0 & 0 & t_1 & t_2  & 0 & 0 & 0\\
          0 & 0 & 0 & 0 & t_1& t_2 & 0 & 0 \\
          0 & 0 & 0 & 0 & 0 & t_1 & t_2 & 0 \\
          0 & 0 & 0 & 0 & 0 & 0 & t_1 & t_2 \\
        \end{array}
      \right)
\end{equation}
and
\begin{equation}\label{HKink}
  \Delta=\left(
        \begin{array}{cccc cccc}
          \Delta_1 & 0 & 0 & 0 & 0 & 0 & 0 & 0\\
          \Delta_2 & \Delta_1 & 0 & 0 & 0 & 0 & 0 & 0 \\
          0 & \Delta_2 & \Delta_1 & 0 & 0 & 0 & 0 & 0 \\
          0 & 0 & \Delta_2 & \Delta_1 & 0 & 0 & 0 & 0 \\
          0 & 0 & 0 & \Delta_2 & \Delta_1  & 0 & 0 & 0\\
          0 & 0 & 0 & 0 & \Delta_2& \Delta_1 & 0 & 0 \\
          0 & 0 & 0 & 0 & 0 & \Delta_2 & \Delta_1 & 0 \\
          0 & 0 & 0 & 0 & 0 & 0 & \Delta_2 & \Delta_1 \\
        \end{array}
      \right),
\end{equation}
if the kink is present only in hopping terms. On the other hand, for a kink in superconducting terms, the matrix $\Delta$  should be replaced by
\begin{equation}\label{HKink}
  \Delta=\left(
        \begin{array}{cccc cccc}
          \Delta_1 & 0 & 0 & 0 & 0 & 0 & 0 & 0\\
          \Delta_2 & \Delta_1 & 0 & 0 & 0 & 0 & 0 & 0 \\
          0 & \Delta_2 & \Delta_1 & 0 & 0 & 0 & 0 & 0 \\
          0 & 0 & \Delta_2 & \Delta_1 & 0 & 0 & 0 & 0 \\
          0 & 0 & 0 & \Delta_1 & \Delta_2  & 0 & 0 & 0\\
          0 & 0 & 0 & 0 & \Delta_1& \Delta_2 & 0 & 0 \\
          0 & 0 & 0 & 0 & 0 & \Delta_1 & \Delta_2 & 0 \\
          0 & 0 & 0 & 0 & 0 & 0 & \Delta_1 & \Delta_2 \\
        \end{array}
      \right).
\end{equation}


\begin{thebibliography}{99}
\bibitem[1]{Kitaev}{A. Y. Kitaev, Sov. Phys.-Usp. 44, 131 (2001).}
\bibitem[2]{Alicea}{J. Alicea, Rep. Prog. Phys. 75, 076501 (2012).}
\bibitem[3]{Leijnse}{M. Leijnse and K. Flensberg, Semicond. Sci. Technol. 27, 124003 (2012).}
\bibitem[4]{Lutchyn}{R. M. Lutchyn, J. D. Sau, and S. Das Sarma, Phys. Rev. Let. {\bf 105}, 77001(2010).}
\bibitem[5]{Oreg}{ Y. Oreg, G. Refael, and F. V. Oppen, Phys. Rev, Let. {\bf 105}, 177002 (2010).}
\bibitem[6]{Sato}{ M. Sato, Y. Takahashi, and S. Fujimoto, Phys. Rev. B {\bf 82}, 134521 (2010).}
\bibitem[7]{Potter}{  A. C. Potter and P. A. Lee, Phys. Rev. Letters {\bf 105}, 227003 (2010).}
\bibitem[8]{Motohiko}{ M. Ezawa Phys. Rev. B, 100, 045407(2019).}
\bibitem[9]{Ning}{N. Wu and W.-L. You Phys. Rev. B 100, 085130( 2019).}
\bibitem[10]{Beenakker}{C. W. J. Beenakker, Annu. Rev. Con. Mat. Phys. 4, 113 (2013).}
\bibitem[11]{Bernevig}{A. Bernevig and T. Neupert, “Topological superconductors and category theory,” arXiv:1506.05805 (2015).}
\bibitem[12]{Altland}{A. Altland and M. R. Zirnbauer, Phys. Rev. B, 55, 1142 (1997).}
\bibitem[13]{Schnyder}{A. P. Schnyder, S. Ryu, A. Furusaki, and A.W. W. Ludwig, Phys. Rev. B, 78, 195125 (2008).}
\bibitem[14]{Sticlet}{D. Sticlet, L. Seabra, F. Pollmann, J. Cayssol, Phys. Rev. B 89, 115430 (2014).}
\bibitem[15]{Yahyavi}{ M. Yahyavi, B. Hetenyi, and B. Tanatar, Phys. Rev. B 100, 064202 (2019).}
\bibitem[16]{Wakatsuki}{R. Wakatsuki, M. Ezawa, Y. Tanaka, and N. Nagaosa, Phys. Rev. B, 90 014505 (2014).}
\bibitem[17]{Xiong}{Y.  Xiong and P.  Q. Tong, New. J. Phys. 17, 013017 (2015).}
\bibitem[18]{Wang}{ Y.  C. Wang, J.  J. Miao, H. K. Jin, and S. Chen, Phys. Rev. B 96, 205428 (2017).}
\bibitem[19]{Ezawa2}{ M. Ezawa, Phys. Rev. B 96, 121105(R) (2017).}
\bibitem[20]{Hua}{ Ch.-B. Hua, R. Chen, D.-H. Xu, and B. Zhou, Phys. Rev. B 100, 205302 (2019).}
\bibitem[21]{Mourik}{ V. Mourik, K. Zuo, S. M. Frolov, S. R. Plissard, E. P. A. M. Bakkers, and L. P. Kouwenhoven, Science 336, 1003 (2012).}
\bibitem[22]{Nadj}{ S. Nadj-Perge et al., Science 346, 602 (2014).}
\bibitem[23]{Tadeu}{ A. Kobialka, N. Sedlmayr, M. M. Maska, T. Domanski, arXiv:1909.11550v1.}
\bibitem[24]{Drost}{  R. Drost, T. Ojanen, A. Harju, and P. Liljeroth, Nature Phys. 13, 668 (2017).}
\bibitem[25]{Leseleuc}{S. Les\'{e}l\'{e}uc et. al., Science 365, 775 (2019).}
\bibitem[26]{Rufo}{S. Rufo, N. Lopes, M. A. Continentino, M. A. R. Griffith, Phys. Rev. B 100, 195432(2019).}
\bibitem[27]{Maffei}{ M.  Maffei, A. Dauphin, F. Cardano, M. Lewenstein, and Massignan, New J. Phys. 20, 013023 (2018).}
\bibitem[28]{Zak}{ J. Zak, phys. Rev. Lett., 62, 23 (1989).}
\bibitem[29]{Fradkin}{E. Fradkin, Field Theories of Condensed Matter Physics, 2nd ed. (Cambridge University Press, New York, 2013).}
\bibitem[30]{Roger}{R. S. K. Mong and V. Shivamoggi, Phys. Rev. B 83, 125109 (2011).}
\bibitem[31]{Jin}{ L. Jin, P. Wang, Z. Song, Sci. Rep. 7, 5903 (2017).}
\bibitem[32]{Continentino}{ M.A. Continentino, H. Caldas, D. Nozadze and N. Trivedi, Phys. Lett. A 378, 3340 (2014).}
\bibitem[33]{Dominguez}{ F. Dominguez, J. Cayao, P. S. Jose, R. Aguado, A. L. Yeyati and E. Prada, npj Quantum Materials 2, 13(2017).}
\end{thebibliography}
\end{document}